\documentclass[showpacs,twocolumn,aps,prx,superscriptaddress,10pt]{revtex4-1}
\usepackage{amsmath,amssymb,amsfonts,bm,graphicx,mathtools,braket}
\usepackage{hyperref}
\hypersetup{%
	pdftitle={dynamical condensate}, %
	pdfauthor={Zhiyuan Sun},%
	pdfpagemode={UseNone},%
	pdfstartview={FitH},%
	breaklinks=true,%
	citecolor=blue,%
	colorlinks=true,%
	linkcolor=blue,%
	urlcolor=blue
}	
\usepackage[T1]{fontenc}
\usepackage{fourier}
\usepackage{baskervald}

\newcommand{\unit}[1]{\,\mathrm{#1}} 
\newcommand{\equa}[1]{Eq.~\eqref{#1}} 
\newcommand{\fig}[1]{Fig.~\ref{#1}}

\newcommand{\rom}[1]{\uppercase\expandafter{\romannumeral #1\relax}}

\graphicspath{{./}}

\begin{document}

\title{Dynamical exciton condensates in biased electron-hole bilayers}

\author{Zhiyuan Sun}
\affiliation{State Key Laboratory of Low-Dimensional Quantum Physics and Department of Physics, Tsinghua
	University, Beijing 100084, P. R. China}
\affiliation{Department of Physics, Columbia University, 538 West 120th Street, New York, New York 10027}

\author{Yuta Murakami}
\affiliation{Center for Emergent Matter Science, RIKEN, Wako, Saitama 351-0198, Japan}

\author{Tatsuya Kaneko}
\affiliation{Department of Physics, Osaka University, Toyonaka, Osaka 560-0043, Japan}

\author{Denis Gole\v{z}}
\affiliation{Jo\v{z}ef Stefan Institute, Jamova 39, SI-1000, Ljubljana, Slovenia}

\author{Andrew J. Millis}
\affiliation{Department of Physics, Columbia University, 538 West 120th Street, New York, New York 10027}
\affiliation{Center for Computational Quantum Physics, Flatiron Institute, 162 5th Avenue, New York, NY, 10010}

\begin{abstract}
Bilayer materials may support interlayer excitons comprised of electrons in one layer and holes in the other. In experiments, a non-zero exciton density is typically sustained by a bias chemical potential, implemented either by optical pumping or by   electrical contacts connected to the two layers. We show that if charge can tunnel between the layers, the chemical potential bias means that an exciton condensate is in the dynamical regime of ac Josephson effect. It has  physical consequences such as tunneling currents and  the ability to tune a condensate from bright (emitting coherent photons) to dark by experimental controlling knobs.
If the system is placed in an optical cavity, coupling with cavity photons favors different dynamical states depending on the bias, realizing superradiant phases. 
\end{abstract}

\maketitle

\begin{figure}[t]
	\includegraphics[width= \linewidth]{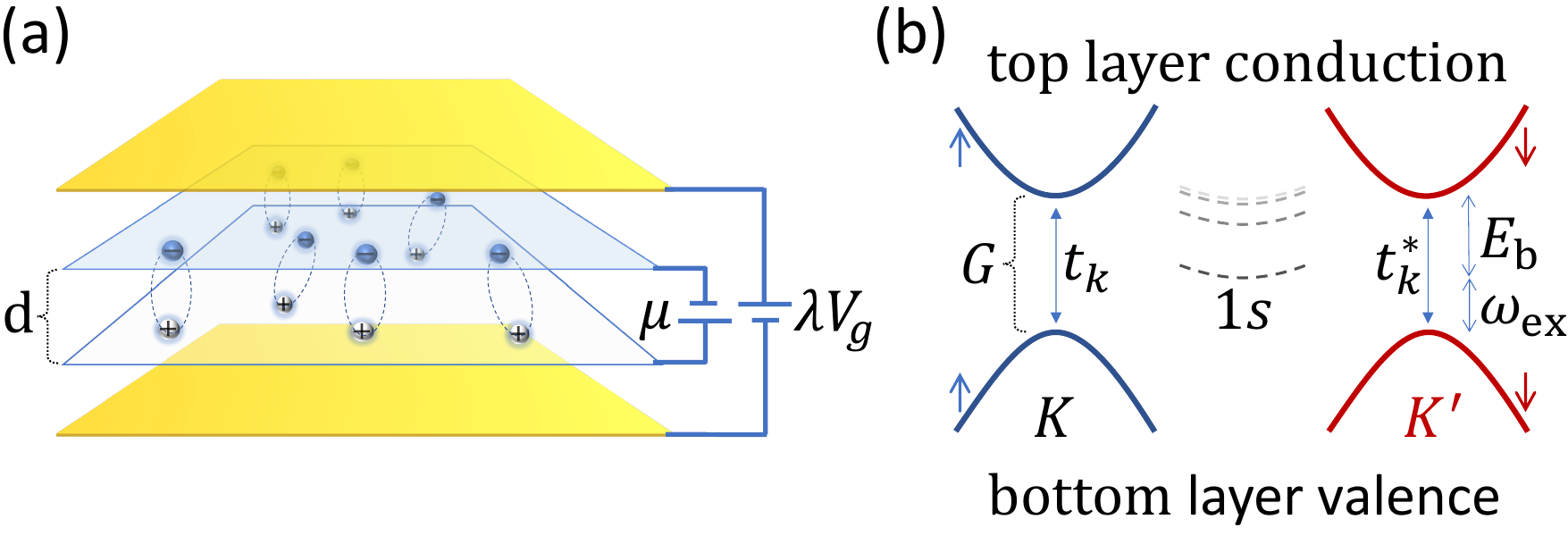} 
	\caption{(a) Schematic of the biased TMDB. (b) The low energy electronic bands giving the interlayer excitons: conduction bands from the top layer and valence bands from the bottom layer. Both bands have   two valleys $K$ and $K^\prime$ with effective mass $m$. The arrows represent the spin eigenvalues. The  dashed curves are schematic exciton energy levels, with the lowest one being the 1s exciton at frequency $\omega_{\text{ex}}$. The gate bias $\lambda V_{\text{g}}$ exerts a vertical electric field that shrinks the original band gap $G_0$  to $G=G_0-V_{\text{g}}$, where $\lambda>1$ because of the thickness mismatch.  The contact bias  applies a chemical potential bias $\mu$ to inject the excitons \cite{Xie2018}.}
	\label{fig:Biased_TMDB}
\end{figure}

An exciton is a boson formed when an electron in a conduction band of an insulator  becomes bound to a hole in its valence band. Interlayer excitons are formed of electrons and holes residing in different layers of a two dimensional (2D) `electron-hole bilayer' system, which are of great current interest for their long life time and because their density, binding energy and other properties may be easily tuned by applied electric fields~\cite{Regan.2022,Rivera2018}.
The energy $\omega_{\text{ex}}$ of an exciton at zero momentum is the band gap $G$ minus the exciton binding energy $E_{\text{b}}$. 
An exciton condensate may be formed in equilibrium at low temperatures if $\omega_{\text{ex}}$ becomes negative due to, e.g., a reduced band gap tuned by  a gate voltage $V_{\text{g}}$ in  \fig{fig:Biased_TMDB}(a)~\cite{Lozovik1976,S.I.Shevchenko1977,Datta.1985,Zhu1995,Naveh.1996_InAs_GaSb,Fogler2014a, Wu2015, Zhu2019}.
Possible equilibrium condensates were studied in bilayers made of InAs/GaSb~\cite{Du2017a} and  transition metal dichalcogenides (TMD)~\cite{, Zhang2021,Gu2021}, and also in the quantum hall regime of bilayers made of GaAs \cite{Spielman2000, Eisenstein2014}, graphene \cite{Li2017,Liu.2017, Liu.2022} and TMD~\cite{Shi2021}.

However, in most experimental systems to date, the exciton condensates are actually non-equilibrium ones  maintained via injection of electrons and holes into the system by optical excitation~\cite{Butov1994,High2012, Sigl.2020_condensate}, or  electrical injection of carriers via conducting leads~\cite{Xie2018,Wang2019d, Ma.2021strongly,Wang.2023_InAs_GaSb,Hori.2023_Si_FET_EI}.
Carrier injection implies a chemical potential difference ($\mu$ in  \fig{fig:Biased_TMDB}(a)) between the layers, which drives a time dependence of the phase of the excitonic order parameter, similar to optically pumped exciton polariton condensates \cite{Wouters.2007, Szymanska.2007,
Deng2010,
Dagvadorj2015,Hanai.2019,
Sieberer.2023_RMP}.  If the electron and  hole layers are electrically isolated from each other,
the order parameter dynamics has no physical effects.
However, we observe that if electrons can tunnel between layers,  it results in an AC Josephson effect~\cite{Sun2021a, Perfetto.2019, Perfetto.2020_pumped_p_exciton, Murakami.2020} that leads to important physical consequences. This includes  coherent photon emission, a DC tunneling current, and remarkable tunability of  phase transitions between bright and dark condensates.  We further show that when placed in an optical cavity~\cite{Shan.2023_MoSe2_cavity, Strashko.2020}, the coupling of the condensate with cavity photons selects different dynamical phases depending on the bias voltage, realizing the long sought super radiant states in non-equilibrium.
We analyze this effect in the context of TMDBs,  but the theory also apply to pumped TMD monolayers and other electron hole bilayers such as GaAs structures, see supplemental information (SI) Sec.~I.

\emph{The device  and the excitonic Hamiltonian---}Biased  TMDB provide a favorable environment for high temperature excitonic condensates~\cite{Fogler2014a, Wu2015, Regan.2022} and recently, exciting experiments have reported their signatures in these systems  \cite{Wang2019d, Ma.2021strongly}. 
The TMDB systems are composed of two layers  of TMD stacked one on top of the other as shown in \fig{fig:Biased_TMDB}(a). 
The hexagonal monolayers are semiconductors with direct bandgaps of $1 -  2 \unit{eV}$ at the $K$ and $K^\prime$ points of the hexagonal Brillouin zone \cite{Rivera2018}. Near $K$ and $K^\prime$, spin-orbit coupling (SOC)  splits the valence band  by about $ 200 \unit{meV}$, and  the conduction band by an amount that varies between $3\sim 60 \unit{meV}$ depending on the specific compound~\cite{Xiao2012,Liu2013a}. 
Applying a gate voltage to  TMDB shrinks the effective interlayer band gap by (say) lowering the energy of the top layer conduction band and raising the energy of the bottom layer valence band, making the interlayer excitons the lowest energy ones \cite{Fogler2014a,Wu2015}. For the low energy excitons, it is enough to  consider only the lowest conduction band from the top layer and the highest valance band from the bottom layer,  as shown in \fig{fig:Biased_TMDB}(b).

\begin{figure}
	\includegraphics[width= \linewidth]{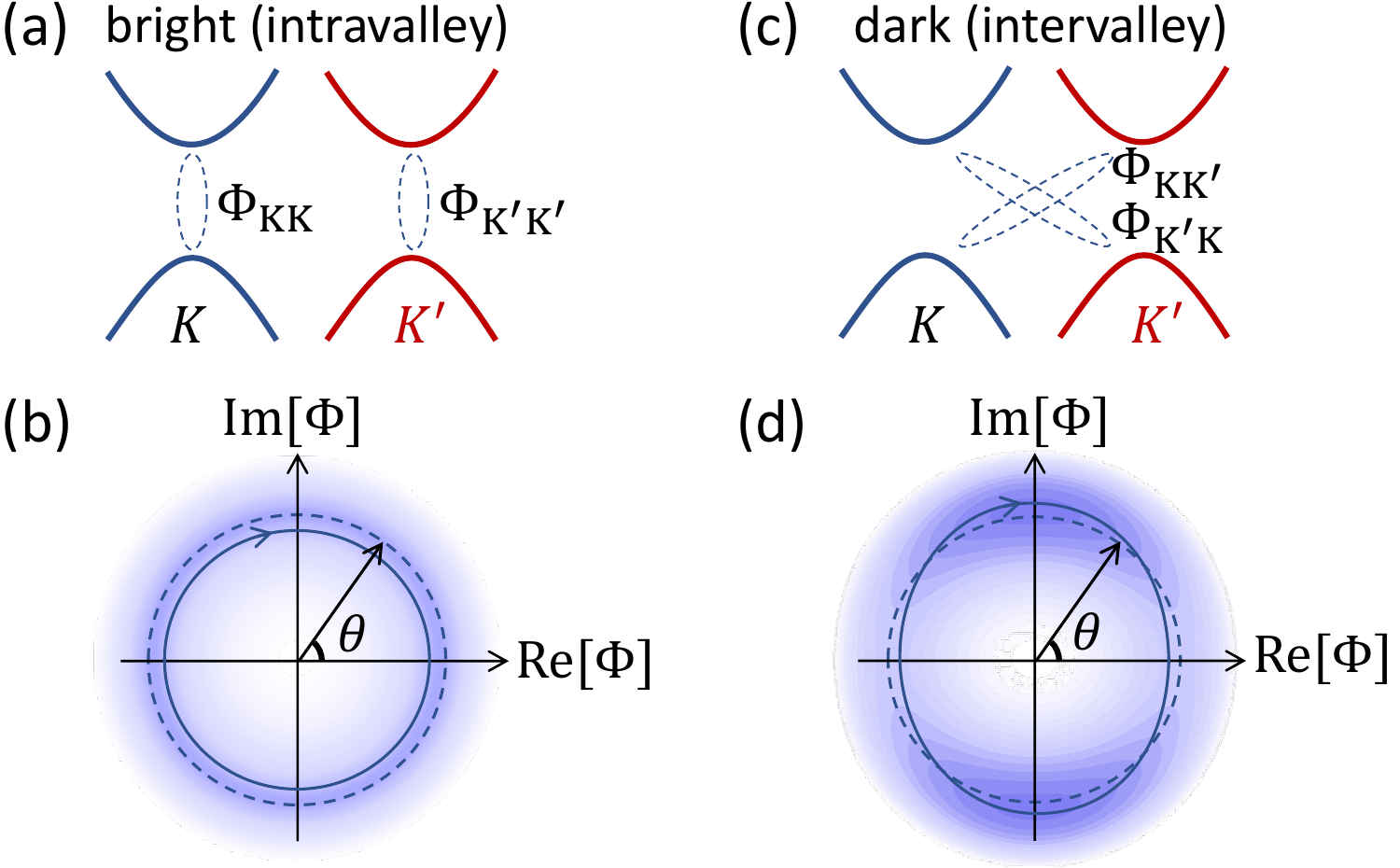} 
	\caption{(a) Illustrations of the bright (intravalley) condensate  with the dashed ovals denoting the electron-hole pairing. 
	(b) The free energy  plotted on the complex plane of the order parameter $|\Phi|e^{i\theta}$  for the bright condensate, with lower energy appearing bluer. The dashed line is the orbit of the dynamical order parameter without interlayer tunneling ($v_{\text{t}}=0$). The solid line is that with tunneling ($v_{\text{t}} \neq 0$). (c,d) Same as (a,b) but for the dark (intervalley) condensate . }
	\label{fig:Static_condensate}
\end{figure}

We focus on the lowest energy s-excitons represented by the bosonic field $\Phi_{ij}$, which  is a $2\times 2 $ matrix  with $i,j$ taking the values of either $K$ or $K^\prime$, giving the four types of excitons: $\Phi_{11}, \Phi_{22}$ means the two intravalley excitons and $\Phi_{12}, \Phi_{21}$ means the two intervalley excitons  (\fig{fig:Static_condensate}). It is convenient to represent it as $\Phi = \Phi_\mu \sigma_\mu/\sqrt{2}=\left(\Phi_0 \sigma_0 + \Phi_1 \sigma_1+ \Phi_2 \sigma_2 + \Phi_3 \sigma_3 \right)/\sqrt{2}$ where $\sigma_\mu$ are the usual Pauli matrices, grouping $\Phi$ into the valley singlet $\Phi_0$ and the valley triplet $\Phi_1, \Phi_2, \Phi_3$.
The effective Lagrangian  is derived as $L = \int d^2 \mathbf{r} 
\left(
\text{Tr} [-i \Phi^\dagger \partial_t   \Phi] + H  
\right)
$ where the Hamiltonian density $H=H_0+H_{\text{J}}$ has two parts:
\begin{widetext}
\begin{align}
	H_0 = &  \text{Tr} \left[ \Phi^\dagger \left(\omega_{\text{ex}}  + \frac{p^2}{4m}
	\right)  \Phi + c_4 (\Phi^\dagger \Phi)^2
	\right]
	+ \frac{1}{2} \int dr dr^\prime \rho(r) U(r-r^\prime) \rho(r^\prime)
	\,,
	\notag\\
	H_{\text{J}} = &
	c_{\text{em}} \left( E_x  \Phi_{0} + iE_y  \Phi_{3} + c.c. \right) +
c_{\text{J}} 
\left[
\frac{1}{2}(\Phi_1^2 + \Phi_2^2+c.c.) +\rho
\right] 
\,,
\label{eqn:exciton_Lagrangian}
\end{align}
\end{widetext}
 see SI Sec.~III.
Note that we set the elementary charge $e$, the Planck constant $\hbar$ and the speed of light $c$ to  unity for notational simplicity, except when we discuss physical observables.
Here $\omega_{\text{ex}}=G-E_{\text{b}}$ is the single exciton energy at zero center-of-mass momentum, 
$p=-i\nabla$,
$\rho=\text{Tr} \left[ \Phi^\dagger \Phi \right]=\Phi_\mu^\ast \Phi^\mu$ is the local exciton density, and $U(r-r^\prime)$ is the kernel for dipole-dipole interaction between excitons~\cite{Wu2015}. For long wavelength behavior of the  condensate, it is enough to approximate $U$  by a local interaction with kernel $U_{0} \equiv \int d^2\mathbf{r} U(r)=4\pi  d/\epsilon$ which is just the inverse of the capacitance $C$ of the bilayer  with the dielectric $\epsilon$, such that the Hartree term becomes $U_0  \int dr \rho^2/2$.
The coefficient $c_4$ is the exchange interaction which is repulsive for adjacent bilayers such that the interlayer distance is small~\cite{Ciuti1998,Combescot2015,Wu2015}.
For an interlayer distance  $d$ much smaller than the excitonic Bohr radius $a_0=2\epsilon \hbar^2/me^2$, one has $c_4 \approx  8\pi /m$, larger than the dipole repulsion $U_0$ by a factor $a_0/d$.

The  $H_{\text{J}}$ part contains the Josephson terms due to the weak interlayer tunneling (hoping) that reduces the $U(1)$ symmetry \cite{Sun2021a,Kaneko2020}. 
We focus on four ($H_M^M$, $H_X^M$, $R_M^X$, $R_M^M$) of the six high symmetry interlayer stackings, where the conduction and valence bands in each valley have different eigenvalues under $C_3$ rotation around certain high symmetry centers~\cite{Tong2017, Rivera2018}, forbidding a direct hoping at the  $K$ and $K^\prime$ points. Together with time reversal 
symmetry, the hoping matrix element is constrained to $t_{\text{k}}=v_{\text{t}} (k_x \pm ik_y)$ for the $K/K^\prime$ valley where $v_{\text{t}} \sim 10^3 - 10^4 \unit{m/s}$ is a velocity scale \cite{Tong2017}. 
This chiral tunneling leads to the second order Josephson term~\cite{Sun2021a} with the Josephson energy $c_{\text{J}}=m v_{\text{t}}^2  f_{\text{J}}(E_{\text{b}}, G, \omega)$  in \equa{eqn:exciton_Lagrangian}, and the  coupling coefficient $c_{\text{em}}= \sqrt{m v_{\text{t}}^2/E_{\text{b}} }
f_{\text{em}}(E_{\text{b}}, G, \omega)$  to the in-plane electric field $\mathbf{E}=-\partial_t \mathbf{A}$, where $f_{\text{J}}$ and $f_{\text{em}}$ are dimensionless functions, see SI Sec.~III. The coupling term to EM field reflects the circular optical selection rules for the bright excitons.

Without the leads,  the classical equation of motion would be $\partial_{\Phi^\dagger} L=0$, or in other words, $i\partial_t \Phi = \partial_{\Phi^\dagger} H$.  The leads are modeled as a bath imposing the chemical potential $\mu$  for the  interlayer excitons. In the simplest case, its only effect is  to add a dissipative term to the equation of motion: 
\begin{align}
i\partial_t \Phi = \partial_{\Phi^\dagger} H -  i\Gamma \partial_{\Phi^\dagger} F  \,, \quad 
F=H-\mu \rho
\label{eqn:EOM}
\end{align}
which acts to drag $\Phi$ to the value that minimizes the  free energy $F$, see Ref.~\cite{Zeng.2023_dynamical_exciton} for a detailed derivation with a bath.
Here $\Gamma$ is the  dimensionless tunneling rate of excitons between the bilayer system and the leads. We assume the inter-layer charge tunneling is the bottleneck of particle flow, such that the exciton tunneling between the system and the leads is effectively uniform  all over the 2D device.

Supplemented by the free Lagrangian~\cite{Sun.2020_superconductor} of the electromagnetic (EM) field $\mathbf{A}$,  Eqs.~\eqref{eqn:exciton_Lagrangian}\eqref{eqn:EOM} describes the device of interest.

\emph{Without interlayer tunnelling ($H_{\text{J}}=0$),} \equa{eqn:exciton_Lagrangian} has a U(2)$\times$U(2) symmetry of valley rotations ($\Phi \rightarrow U_1^\dagger \Phi  U_2$ where $U_1, U_2$ are $2\times2$  unitary matrices) \cite{Wu2015}, same as the excitons made of spin $1/2$ electrons and holes. The steady state is an `equilibrium state' which simply minimizes the free energy $F$~\cite{Murakami:2022aa}. For $\mu>\omega_{\text{ex}}$ and well below  the  Berezinskii-Kosterlitz-Thouless (BKT) temperature~\cite{Berezinsky1971,Kosterlitz1973}, an exciton condensate occurs  and the bosonic field  develops a classical value $\Phi$: the order parameter.
Note that the repulsive exchange interaction 
\begin{align}
	\text{Tr}[(\Phi^\dagger \Phi)^2]=\frac{1}{2}\rho^2 +\frac{1}{2}\sum_{l=1,2,3} \left( \Phi_0^\ast \Phi_l + c.c. + i\epsilon_{ljk} \Phi_j^\ast \Phi_k \right)^2 
\end{align} 
favors the states with equally distributed exciton numbers between the two valleys. Therefore, the typical lowest energy condensate satisfies $|\Phi_{\text{KK}}|=|\Phi_{\mathrm{K^\prime K^\prime }}|$ and  $|\Phi_{\mathrm{K K^\prime }}|=|\Phi_{\mathrm{K^\prime K}}|$  such that the most generic ground state order parameter can be written as 
\begin{align}
	& \Phi_\mu =e^{i\theta}(i a_0, a_1, a_2, a_3),\quad 
		a_\mu a^\mu =\rho_0=\frac{\mu-\omega_{\text{ex}}}{2g}
	\,
	\label{eqn:S1_S3}
\end{align}
where $a_\mu$ are real numbers and we defined $g=(c_4+U_0)/2$.
Therefore, the ground state manifold is four dimensional with the topology of $S_1 \times S_3$ where $S_{\text{N}}$ means the $N$ dimensional sphere. Choosing one point on this manifold corresponds to spontaneously breaking the continuous symmetries associated with this manifold and implies four gapless Goldstone modes.   The solution of \equa{eqn:EOM} says the overall phase constantly  winds in time: 
\begin{align}
\theta=\theta(0)- \mu t \equiv \theta_0(t)
	\label{eqn:phase}
\end{align}
meaning the state may be represented as an one dimensional orbit in the order parameter space, as shown by the dashed curves in  \fig{fig:Static_condensate}(b)(d). However, this dynamics doesn't have any observable effects.


\emph{Effect of interlayer tunneling---}We now turn to the central topic of this paper: a nonzero electronic interlayer tunneling $v_{\text{t}}$ results in $H_{\text{J}}$ in \equa{eqn:exciton_Lagrangian} and renders the dynamical condensate a truly non-equilibrium steady state. 
From the symmetry point of view, $H_{\text{J}}$ breaks the conservation of exciton number and reduces  the  valley rotation symmetry to O(2)$\times$O(2) rotations  around $\sigma_3$ only.  Two basic candidates of dynamical `ground states' are  the bright (intravalley, \fig{fig:Static_condensate}(a)(b)) and dark (intervalley, \fig{fig:Static_condensate}(b)(d)) condensates which have different properties.

The bright condensate (\fig{fig:Static_condensate}(a)(b)) corresponds to nonzero $\Phi_0$ or $\Phi_3$ or their linear combination. From $H_{\text{J}}$, this state has an in-plane electrical polarization $(P_x, P_y)= 2 c_{\text{em}}  (-\text{Re}[\Phi_0], \text{Im}[\Phi_3])$, which is dynamical since it is locked with the winding order parameter phase $\theta(t)$. For example, if only $\Phi_3=\sqrt{\rho} e^{i\theta(t)}$ is nonzero, the polarization oscillates along the $y$-direction and  emits  coherent EM wave that propagates vertically away from the 2D device with the electric field $E_y=\frac{2\pi}{c} \partial_t P_y$ and radiation power
\begin{align}
	P_{\text{r}}&=\frac{c}{4\pi}E_y^2 = 2 \alpha \frac{m v_{\text{t}} ^2 E_{\text{b}}}{G^2} \mu^2 \rho \equiv \gamma_r \mu \rho
	,
	\label{eqn:radiation_power}
\end{align}
as the name `bright' suggests.  Here $\gamma_r$ is the  radiative decay rate of intravalley excitons, $E_{\text{b}}$ is the binding energy of the 1s exciton, $\alpha=e^2/(\hbar c) \approx 1/137$ is the fine structure constant and we have temporarily restored $e$ and $\hbar$. The radiation power also implies a tunneling charge current: $I_{\text{bright}}=P_{\text{r}}/\mu=\gamma_r \rho$.   We have neglected dielectric screening $\epsilon$ since it does not affect such long wavelength physics for typical thin devices.
Note that the polarization of emitted photons depends continuously on the ratio between $\Phi_0$ and $\Phi_3$, which is spontaneously chosen by the non-equilibrium symmetry-broken state.

The dark condensate (\fig{fig:Static_condensate}(c)(d)), corresponding to nonzero $\Phi_1$ or $\Phi_2$ or their linear combination, does not emit photons, as its name suggests. Nevertheless, it experiences a second order Josephson effect~\cite{Sun2021a}. Taking $\Phi_1$ as an example, the free energy landscape plotted on the complex order parameter plane is no longer invariant in the phase direction, but is distorted by the weak Josephson term $c_{\text{J}}  \rho \cos (2\theta)$  in $H_{\text{J}}$ of \equa{eqn:exciton_Lagrangian}, as shown by \fig{fig:Static_condensate}(d). Consequently, the system is in the AC Josephson effect regime with an interlayer charge current $J=c_{\text{J}} \rho \sin (2\theta)$ oscillating at the frequency $2\mu$, and a measurable DC current which we will discuss later.

\begin{figure*}
	\includegraphics[width= \linewidth]{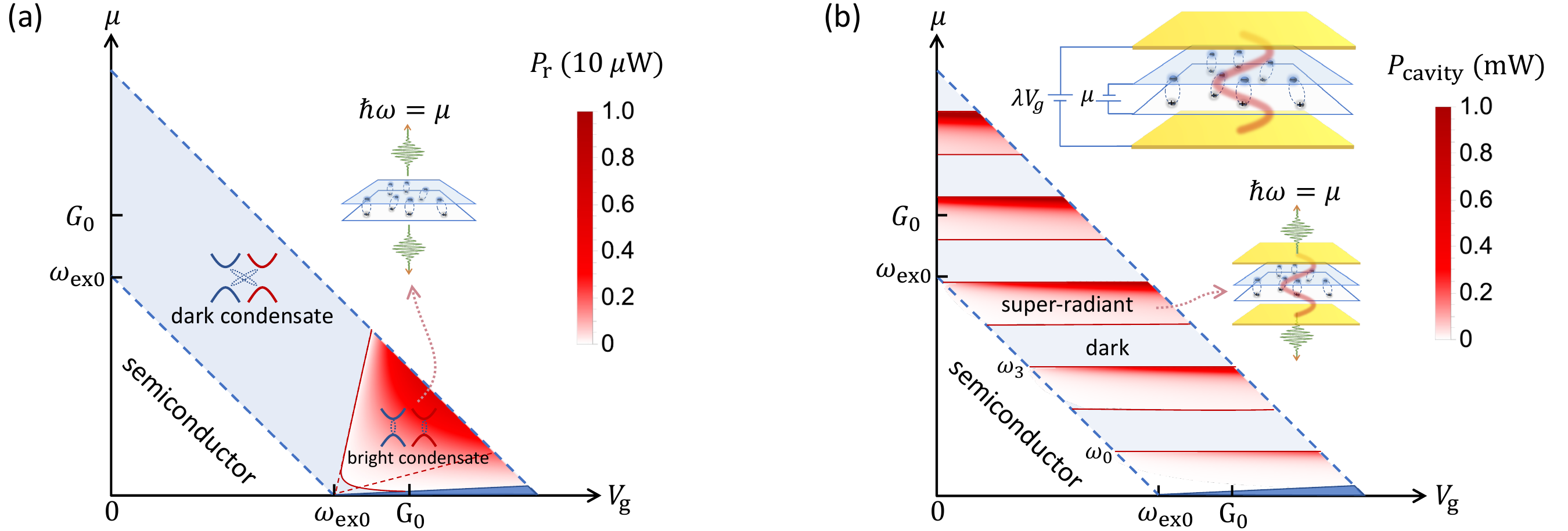} 
	\caption{
		(a) The phase diagram of the non-equilibrium steady states on the $\mu-V_{\text{g}}$ plane of the device (\fig{fig:Biased_TMDB})  well below the BKT temperature. The colored area is the exciton condensate phase, with the light blue/red region being the dark/bright dynamical condensate, and the dark blue region being the dark static  condensate. The color scale in red is the coherent photon radiation power $P_{\text{r}}$ of a $1 \unit{\mu m}^2$ device, also meaning a tunneling current $P_{\text{r}}/\mu$. When computing $P_{\text{r}}$, we have assumed $f_{\text{em}}(E_{\text{b}}, G, \omega)=f_{\text{em}}(E_{\text{b}}, G_0, \omega_{\text{ex}})$ since this function does not change dramatically across the phase diagram of interest.
		The other parameters are $G_0=1.50 \unit{eV}$, $\omega_{\text{ex0}}=G_0-E_{\text{b}}=1.23 \unit{eV}$, $\gamma=0.10 \unit{eV}$ and the  typical ones as in table~I of the SI.
		(b) Same as (a) but for the device in an optical cavity shown by the top inset, where the red wavy curve  represents a cavity photon mode driven by the condensate.  The cavity  thickness is $2h=0.8 \unit{\mu m}$, the dielectric is $\epsilon=10$, and the radiative damping rate of cavity modes is $\gamma_{\text{p}}=30 \unit{meV}$. The red region now means the `superradiant' bright condensate.
	}
	\label{fig:Dynamic_condensate}
\end{figure*}

\emph{The non-equilibrium phase diagram---}The remaining question is  which dynamical state the system prefers. In this paper, we address this question at the mean field level, meaning we neglect spatial fluctuations, and construct an effective static potential which the dynamical state should minimize.  
Without loss of generality, we pick $\Phi_3$ for the bright condensate and  $\Phi_1$ for the dark one, and write the order parameter as somewhere between them: 
\begin{align}
(\Phi_1, \Phi_3)=  (\sqrt{\eta}, \sqrt{1-\eta}e^{i\theta_a}) \sqrt{\rho} e^{i\theta}
\,.
\label{eqn:eta}
\end{align}
Here $\eta \in [0,1]$  controls which state the system lies in: $\eta=0$ means the bright condensate and $\eta=1$ means the dark one. \equa{eqn:EOM} could be rewritten for the new variables $\rho$, $\theta$, $\theta_a$ and $\eta$:
\begin{align}
&\dot{\theta} =- \partial_{\rho} H ,\quad
\dot{\rho} = \partial_{\theta} H - 4 \Gamma \rho \partial_{\rho} F ,\quad
\notag\\ 
& \rho \dot{\theta}_a =\partial_{\eta} H  ,\quad
\partial_t\left[ (1-\eta) \rho \right] = \partial_{\theta_a} H  
\,.
\label{eqn:EOM_eta}
\end{align}
For simplicity, we kept only the  amplitude ($\rho$) direction of the damping term $\Gamma \partial_{\Phi^\dagger} F$, which does not change the qualitative conclusion.
In the limit of $H_{\text{J}}=0$, the solution is simply an orbit $S_0=\left(\theta_0(t), \rho_0, 0, \eta_0 \right)$ with $\theta_0(t)$ defined in \equa{eqn:phase}, a constant  exciton density $\rho=\rho_0$ defined in \equa{eqn:S1_S3},  $\theta_a=0$, and a spontaneously chosen $\eta_0$. 

A weak $H_{\text{J}}$ distorts the orbit, which could be written  perturbatively in $v_{\text{t}}$:
\begin{align}
(\theta, \rho, {\theta}_a, \eta)=
S_0+S_1
+ O(c_{\text{J}}^2)
	\,.
	\label{eqn:orbit}
\end{align}
where $S_1=\left(\theta^{(1)}, \rho^{(1)}, {\theta}^{(1)}_a, \eta^{(1)} \right)$ is the $O(c_{\text{J}})$ correction to the orbit.
In the dark condensate ($\eta=1$), all the variables oscillate  around the perfect circle $S_0$ with frequency $2\mu$,
 as  shown  schematically by the solid curve in \fig{fig:Static_condensate}(d). In the bright condensate ($\eta=0$), the photon radiation tends to reduce the exciton density and drags the orbit smaller, as shown by the solid curve in \fig{fig:Static_condensate}(b). Note that the actual bright orbit also oscillates due to the linear polarization of the coherent emission, which for simplicity is not shown. 

This fast oscillation exerts  a slow force~\cite{Sun.2023_ponderomotive, Wan.2017_Frustrated_Magnet_floquet} on the slow component of $\eta$, the Pondermotive force~\cite{Sun.2023_ponderomotive}, which could be computed by time averaging the total force for $\eta$:
\begin{align}
F_{\text{P}} &= \langle - \partial_\eta L\rangle_t 
\notag\\
&=  
-\langle 
c_{\text{J}}  \rho \cos 2\theta - \rho \dot{\theta}_a
+ c_4  \rho^2 (1-2\eta) \sin^2 \theta_a
\rangle_t 
\notag\\
&=  
c_{\text{J}}^2  \rho_0
\left[ 
\frac{ 2\mu-4g \rho_0 \eta  + (1-\eta)\frac{\gamma^2}{2\mu} }{4\mu^2 + \gamma^2}
+ c_4 \rho_0 \frac{2\eta-1}{8\mu^2} 
\right]
	\,
	\label{eqn:p_force}
\end{align}
where  the damping rate is $\gamma=8g \rho_0 \Gamma$.
Note that we have neglected the $c_{\text{em}}$ term in $H_{\text{J}}$, whose effect is suppressed by a factor of $\alpha^2 \mu^2/\gamma^2$.
The Pondermotive force corresponds to an effective potential for $\eta$, coined the Pondermotive potential~\cite{Sun.2023_ponderomotive}: $V_{\text{P}} = - \int d\eta  F_{\text{P}}$.


Naturally, dissipation effects from the environment would relax $\eta$ to minimize $V_{\text{P}}$ (this effect is not  contained in \equa{eqn:EOM}), giving the non-equilibrium phase diagram in \fig{fig:Dynamic_condensate}(a). The red region means in general a linear combination between the bright condensate and the dark one ($0 \leq \eta < 1$), with $\eta$ given by $F_{\text{P}}(\eta)=0$. The power of coherent radiation computed from \equa{eqn:radiation_power} is also shown by the red color scale, which implies a DC tunneling current.
The dark condensate (blue region, $\eta=1$) occupies the most experimentally accessible part of the phase diagram. It does not radiate but still has a DC tunneling current 
\begin{align}
	I_{\text{dark}} &= \langle c_{\text{J}} \rho \sin (2\theta) \rangle_t
=
\Gamma \frac{c_{\text{J}}^2}{\mu}  \rho_0 
\frac{
	\mu^2 + \omega_{\text{ex}}^2 + \frac{1}{16} \gamma^2
}
{
	\mu^2+ \frac{1}{16}\gamma^2
}
\,
\label{eqn:dark_current}
\end{align}
which is at the order of $0.1 \unit{\mu A/\mu m^2}$ for typical parameters of TMDB, see SI Sec.~VII. Therefore, one may tune the dynamical condensate across different phases by the gate and bias voltages, with the photon radiation and tunneling currents as their signatures.

\emph{Dynamical condensate in a cavity---}Since the bright condensate radiates, placing the device in a Fabry–Pérot cavity would result in interesting interplay between the condensate and cavity photons~\cite{Schlawin.2022, Regan.2022}, forming exciton-polariton condensates \cite{Deng2010, Shan.2023_MoSe2_cavity}.  

In a cavity of thickness $2h$ formed with walls made of perfect metal and filled with dielectric spacers of dielectric constant $\epsilon$, the cavity photon modes at zero in-plane momentum are polarized in-plane, have frequencies $\omega_{\text{n}}=(n+1)\omega_0$  with indices $n=0,1,2...$  and $\omega_0=c \pi /(2\sqrt{\epsilon}h)$ being the fundamental mode frequency. For simplicity, we assume that the TMDB is placed in the middle of the cavity.  The oscillating in-plane polarization $P$ would linearly drive the discrete photon modes of the cavity. The latter feed back by an oscillating electric field $E=\frac{2\pi}{\sqrt{\epsilon}c} \mu P \tan \frac{\pi \mu}{2\omega_0}$ on the TMDB plane that could be obtained by solving the Maxwell's equations. This electric field couples to the order parameter through the $c_{\text{em}}$ term in \equa{eqn:exciton_Lagrangian}, and would therefore contribute an effective  potential $V_{\text{Pcavity}}$ for the bright condensate relative to the dark one, in the same spirit as that behind the formation of cold atomic super-radiant states~\cite{Kollar.2017_superradiance_cavity}. Rigorously speaking, this effective potential is the Ponderomotive potential~\cite{Sun.2023_ponderomotive} obtained by integrating out the cavity photon modes in the Keldysh path integral, which in the current case is proved to be simply one half of the coupling energy: $V_{\text{Pcavity}}=-\langle P E \rangle_t /2$, giving 
\begin{align}
V_{\text{Pcavity}}= 
- \frac{4 \alpha }{\sqrt{\epsilon} }   
\frac{ \mu E_{\text{b}}}{G^2}  m v_{\text{t}}^2 \rho_0
\mathrm{Re}\left[
\tan \frac{\mu+i\gamma_{\text{p}}}{2 \omega_0 / \pi}
\right]
	\,
\label{eqn:f_cavity}
\end{align}
where we have added the damping rate $\gamma_{\text{p}}$ of the cavity photons.

Since $V_{\text{Pcavity}}$ is at the order of $v_{\text{t}}^2$, it dominates over the $O(v_{\text{t}}^4)$ potential from \equa{eqn:p_force} due to Josephson oscillations. Furthermore, $V_{\text{Pcavity}}$ is negative and resonantly enhanced whenever the dynamical frequency $\mu$ is a little below the frequency $\omega_{\text{n}}$ of an even index cavity mode whose anti-node is on the TMDB, strongly favoring the bright condensate. This is because this cavity mode is driven  almost resonantly and in phase with the polarization of the condensate.  As a result, the phase diagram of this device is modified by the cavity to that in \fig{fig:Dynamic_condensate}(b).

The driving of cavity modes leads to the dissipation power $P_{\text{cavity}}= \langle E \dot{P} \rangle 
= \frac{1}{2\sqrt{\epsilon} } P_{\text{r}} \mathrm{Im}
\left[
\tan \frac{\mu+i\gamma_{\text{p}}}{2 \omega_0 / \pi}
\right]$ with $P_{\text{r}}$ from \equa{eqn:radiation_power}, 
which in turn means a tunneling current $I=P_{\text{cavity}}/\mu$.   
If the damping of cavity photons comes only from leakage out of the cavity, all the power is converted to coherent photon radiation as plotted by the red color scale in \fig{fig:Dynamic_condensate}(b).
Furthermore, when a resonance is met at $\mu \approx \omega_{\text{n}}$, the power of radiation  peaks at $P_{\text{r}} \omega_0/(\sqrt{\epsilon}  \gamma_{\text{p}})$, enhanced by a photon quality factor compared to a bright condensate without cavity. This  realizes a long sought super-radiant state \cite{Littlewood1996} in a solid state system, which is forbidden in equilibrium~\cite{Birula1979, Nataf2010, Andolina2019}.
Therefore, this device may work as a laser~\cite{Schneider.2013,Bhattacharya.2013,Gu.2019,Suchomel.2020,Kavokin.2022, Bloch2022} that converts DC current into coherent photons at the energy of the bias voltage, with the allowed `shining' frequencies tunable by the thickness of the device. 

\emph{Discussion---}The dynamical condensates may be viewed as continuous time crystals: they spontaneously break the time translational symmetry by picking a $\theta$, which is observable by the phase of the coherent emission or the oscillating current~\cite{Wilczek.2013.time.crystal,Watanabe.2015.time.crystal,Khemani.2016,Moessner:2017ur,Zaletel.2023_time_crystal_review}. This dynamics also periodically modulate the dispersion of the Goldstone modes and lead to parametric generation of quantum entangled phasons~\cite{Sun.2022_entangled_plasmon}, see SI Sec.~IX.

The bright condensate may be favored by other mechanisms not explicitly considered here, e.g., a nonzero first order Josephson tunneling, a coupling to lattice distortion,  or by driving the system with coherent light.
We note that a nonzero local electron-hole exchange interaction~\cite{Sethi.2021} that favors the dark condensate (see SI Sec.~I),  the indirect band gaps in some compounds \cite{Hsu2017}, and bi-excitons effects in monolayer TMDs \cite{You2015,Wang2018} may also change the phase diagrams in \fig{fig:Dynamic_condensate}.

It is a meaningful future direction to investigate the fluctuation effects of the dynamical condensate using, e.g.,  numerical methods~\cite{Murakami.2020}. In the case of $H_{\text{J}}=0$ so that the ground state manifold is  $S_1 \times S_3$, previous study of spinor BEC~\cite{Stamper-Kurn.2013} indicates that the thermal phase transition is  of the BKT type~\cite{Berezinsky1971,Kosterlitz1973} controlled by the vortices of the overall phase $\theta$. In the low temperature phase, the $\theta$ field has quasi long range (algebraic) order   while the $S_3$ degrees of freedom are disordered. With nonzero $H_{\text{J}}$, the symmetry is further reduced and quasi long range order probably still exists.   It is also interesting to generalize the current study to TMDBs  with first order Josephson tunneling (SI Sec.~XI), to those with periodic Moire patterns~\cite{Wu2018, Bai.2020_exciton, Kennes.2021_Moire}, and to  double layers in the quantum hall regime~\cite{Li2017,Liu.2017, Liu.2022} .

\begin{acknowledgements}
	Z.S. and A.J.M. acknowledge support from the
	Energy Frontier Research Center on Programmable Quantum Materials funded
	by the US Department of Energy (DOE), Office of Science, Basic Energy
	Sciences (BES), under award No. DE-SC0019443.
	Z. S.  acknowledges the support by the National Key Research and Development Program of China (2022YFA1204700), and the State Key Laboratory of Low-Dimensional Quantum Physics at Tsinghua University. Y.M. and T.K. are supported by Grants-in-Aid for Scientific Research from
	JSPS, KAKENHI Grants No. JP20K14412 (Y.M.), No. JP21H05017 (Y.M.), No.
	JP20H01849 (T.K.), and JST CREST Grant No. JPMJCR1901 (Y.M.).
	We thank L. Ma, S. Zhang, M. M. Fogler, D. N. Basov,  J. Shan,   S. Xu, F. Liu and F. Xuan for helpful discussions.
\end{acknowledgements}

\bibliographystyle{apsrev4-1}
\bibliography{Excitons.bib,Excitonic_Insulator.bib,Excitonic_Insulator_2.bib,Cavity.bib,Nonequilibrium.bib,Parametric_Amplification.bib,My_publication.bib}

\pagebreak
\widetext
\begin{center}
	\textbf{\large Supplemental Information for `Dynamical exciton condensates in biased electron-hole bilayers'}
\end{center}
\setcounter{equation}{0}
\setcounter{figure}{0}
\setcounter{table}{0}
\setcounter{page}{1}
\makeatletter
\renewcommand{\theequation}{S\arabic{equation}}
\renewcommand{\thefigure}{S\arabic{figure}}
\renewcommand{\bibnumfmt}[1]{[S#1]}

\tableofcontents

\section{The devices}
\begin{figure}
\includegraphics[width=\linewidth]{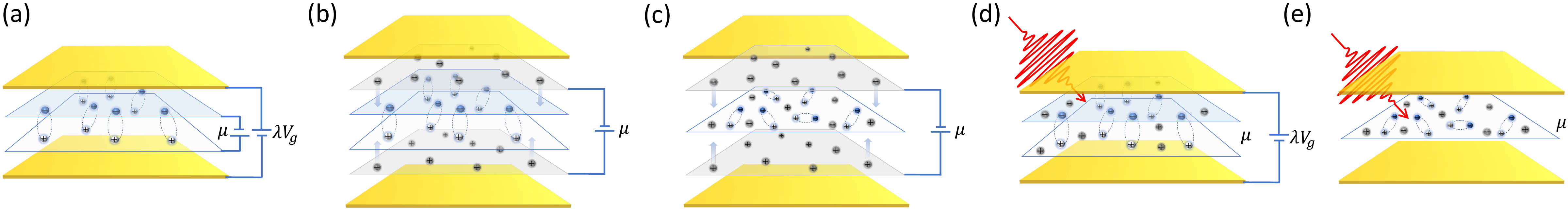} 
\caption{The five types of devices that the theory (Eqs.~\eqref{eqn:exciton_Lagrangian}\eqref{eqn:EOM}) and the qualitative  conclusions apply to. 
(a)The  device that the main text focused on. The excitons are sustained by a chemical potential $\mu$ provided by the  electrical contacts~\cite{Wang.2019,Ma.2021strongly}. The top and bottom gates may play two roles: providing a z-direction electric field that tunes the interlayer band gap, and forming an optical cavity.
(b) In this device, the chemical potential $\mu$ for the excitons is provided by the two tunneling gates~\cite{Xie2018} made of, e.g., graphene. The yellow metallic gates play the role of optical cavity.
(c) Same as (b) but for monolayer excitons.
(d) The reservoir for the excitons is the optically pumped free electrons and holes, which effectively exerts a chemical potential $\mu$. The top and bottom gates play two roles: tuning the gap and forming the cavity.
(e) The excitons are intralayer excitons in a monolayer~\cite{Shan.2023_MoSe2_cavity}. The optically pumped free electrons and holes provide the reservoir  which exerts a  chemical potential $\mu$ for the excitons.
}
\label{fig:Devices}
\end{figure}

Although we used the electrically biased TMD bilayer as an example in the figures of the main text, our theory applies to the optically pumped case equally well, where the bath becomes the free electrons and holes pumped by light~\cite{Butov1994,High2012, Littlewood1996, Szymanska.2007,Hanai.2019}. It also applies to other materials such as graphene  in the quantum hall regime~\cite{Li2017,Liu.2017, Liu.2022} and GaAs~\cite{Spielman2000, Eisenstein2014}.
In \fig{fig:Devices},  we show
five types of devices that the theory and the qualitative  conclusions apply to. 

\section{The electronic Hamiltonian}
Since we are interested in the lowest energy exciton condensates, we keep the four relevant electronic bands: a highest valence band and a lowest conduction band at each valley, represented by the four component electron annihilation operator $\psi(r) = \left( \psi_{K,c}, \psi_{K^\prime,c}, \psi_{K,v}, \psi_{K^\prime,v}  \right)^T$. Here $ \psi_{K,c/v}$ is the annihilation operator of conduction/valence band electrons at the $K$ valley and $ \psi_{K^\prime,c/v}$ is that at the $K^\prime$ valley. There is no spin index since the spins are locked with the valley by the spin orbit interaction, see Fig.~1b of the main text.
The four band Hamiltonian reads
\begin{align}
	& 	H = \psi^\dagger
	\begin{pmatrix}
		\xi_1(p) & \hat{t}_p
		\\
		\hat{t}_p^\dagger & -\xi_2(p)
	\end{pmatrix}
	\psi
	+V_{\text{ee}}+V_{\text{hh}}+V_{\text{eh}}+V_{\text{ex}}
	,
\notag \\
& V_{\text{ee/hh}} =\int dr dr^\prime \frac{1}{|r-r^\prime|} \rho_{\text{e/h}}(r) \rho_{\text{e/h}}(r^\prime)
,\quad
V_{\text{eh}}=-\int dr dr^\prime  \frac{1}{\sqrt{|r-r^\prime|^2+d^2}} \rho_{\text{e}}(r) \rho_{\text{h}}(r^\prime)
,\notag \\
&V_{\text{ex}} = V_{\text{ex0}} \int dr  \sum_{\tau} 
\psi_{\tau, c}^\dagger(r) \psi_{\tau, v}(r) \psi_{\tau, v}^\dagger(r) \psi_{\tau, c}(r) 
,\notag \\
&  \rho_{\text{e}}(r)=\sum_{\tau} \psi_{\tau, c}^\dagger(r) \psi_{\tau, c}(r) 
,\quad
\rho_{\text{h}}(r)=\sum_{\tau}  \psi_{\tau, v}(r)  \psi_{\tau, v}^\dagger(r)
\label{eqn:Hamiltonian}
\end{align} 
where 
$p$ should be understood as a spatial derivative $-i\nabla$ acting on the field operators. The electromagnetic (EM) field represented by the vector potential $A$ enters via the minimal coupling $p\rightarrow p-A$.  
We use quadratic approximation to the band dispersion  $\xi_1(k)=\xi_2(k)=k^2/(2m) + G_0/2-V_{\text{g}}/2 =k^2/(2m) + G/2 = \varepsilon_k + G/2$ with effective mass $m$.
$\hat{t}_p$ is the interlayer hopping. $V_{\text{ee}}$, $V_{\text{hh}}$ and $V_{\text{eh}}$ are the Coulomb interaction between the carriers and  $\rho_{\text{e}}$, $\rho_{\text{h}}$ are the density operators of the electrons and holes. The summation of $\tau$ is over  $K, K^\prime$.
The effective mass has been chosen to be the same for electron and holes for notational simplicity which does not affect the conclusions.

We now discuss the electron-hole exchange interactions~\cite{Yu2014,Xuan.2020_valley_Zeeman, Sethi.2021}. In the case of interlayer excitons, since this exchange requires nonzero overlap between the atomic orbitals of the electron and the hole that reside in different layers, it is much weaker than that of the intralayer excitons. The first exchange interaction is the  weak long range Coulomb interaction between interband transition dipoles (this is not to be confused with the direct dipole-dipole interaction between the permanent dipoles). It is implicitly included by the minimal coupling to the EM field in $\hat{t}_{p-A}$. Together with the Lagrangian of the EM field in vacuum, this coupling leads to the electron-hole `exchange' interaction that splits longitudinal and transverse bright excitons (also called exciton polaritons) at nonzero center of mass momentum $q$~\cite{Yu2014}. However, the splitting is zero at zero $q$ for two dimensions, and is therefore inconsequential for our discussion of the mean field condensate.
The second electron-hole  exchange interaction is the local exchange between overlapping atomic orbitals. In the local limit, this interaction does not distinguish the $K$ and $K^\prime$ valleys, and therefore  does not break the degeneracy between all four types of the lowest energy excitons. 
Away from the local limit, the exchange is represented as the $V_{\text{ex}}$ term in \equa{eqn:Hamiltonian}. 
Since the $C_3$ eigenvalues of conduction and valence electrons at the $K$ valley differ by $e^{i2\pi/3}$, while those at the $K^\prime$ valley differ by $e^{-i2\pi/3}$, the $C_3$ rotational symmetry constrains the two electron-hole pairs of the exchange term to be in the same valley. This means that  this exchange does not have valley rotational symmetry like that for spins. It raises the energy of the bright excitons $\Phi_0, \Phi_3$ by the same amount, although $\Phi_0$ is a valley singlet and $\Phi_3$ belongs to the valley triplet. Since the exchange interaction $V_{\text{ex}}$  is weak and unknown, we don't include it  in the main text.

The interlayer tunneling matrix element $\hat{t}_p$ has contributions from both direct hopping and higher order processes involving remote bands, but its magnitude is small and its  form is controlled by symmetry depending on the way in which the two components of the bilayer are stacked. 
In four of the six high symmetry interlayer stackings 
 ($H_M^M$, $H_X^M$, $R_M^X$, $R_M^M$),
the conduction and valence bands in each valley have different eigenvalues under $C_3$ rotation around certain high symmetry centers, which forbids a direct tunneling at the  $K$ and $K^\prime$ points~\cite{Rivera2018}. Together with time reversal 
($\hat{T}: \left( \psi_{K,c}, \psi_{K^\prime,c}, \psi_{K,v}, \psi_{K^\prime,v}  \right) 
\rightarrow 
\left(\psi_{K^\prime,c},  \psi_{K,c}, -\psi_{K^\prime,v}, -\psi_{K, v}  \right) $) 
symmetry,  the leading terms of the tunneling matrix element is constrained to be \cite{Tong2017}
\begin{align}
	\hat{t}_k= v_{\text{t}} \begin{pmatrix}
		k_x+ik_y  & 0
		\\
		0 & k_x-ik_y  
	\end{pmatrix}
	\label{eqn:tk}
	\,
\end{align}
where $k$ is the crystal momentum measured from the $K/K^\prime$ point, and $v_{\text{t}} \sim 10^3 - 10^4 \unit{m/s}$ for bilayers with no interlayer spacers~\cite{Tong2017}. This form of $t_{\text{k}}$
leads to a second order Josephson effect \cite{Sun2021a} which we will focus on for most of this paper. The other two stacking types ($R^M_X$ and $H^X_X$) will lead to a $t_{\text{k}}$ that is non-vanishing as $k\rightarrow 0$, producing a first order Josephson effect whose consequences are discussed in Sec.~\ref{sec:first_order}.

\section{The effective action for the excitons}
In this section, we  show how to derive the effective action for the excitons from the Hubbard-Stratnovich transformation with the suitable basis (wave functions) for the excitons. For notational simplicity, we  temporarily neglect the interlayer tunneling $\hat{t}$, the electron-electron, hole-hole repulsion $V_{\text{ee}}, V_{\text{hh}}$ and the electron-hole exchange $V_{\text{ex}}$, and add them back latter. The only remaining interaction is the electron hole attraction $V_{\text{eh}}$, which can be decomposed generally with the  Hubbard-Stratnovich field $\Delta$:
\begin{align}
	L_{e} &=  \sum_{k} \psi^\dagger_{k}
	\begin{pmatrix}
		-i\partial_t +	\xi(k) & 0
		\\
		0 & -i\partial_t -\xi(k)
	\end{pmatrix}
	\psi_k
	+
	 \sum_{k,q} \left( \psi^\dagger_{k+q/2} \hat{\Delta}_{k,q}	\psi_{k-q/2}+ c.c. \right)
	+ \sum_{k_1, k_2, q} \text{Tr}[ \hat{\Delta}_{k_1,q}^\dagger 
	V^{-1}_{k_1,k_2} \hat{\Delta}_{k_2,q} ]
\end{align}
where $q$ has the meaning of center of mass momentum of excitons, $V^{-1}_{k_1,k_2}$ is the $k_1, k_2$ component of the inverse of the electron hole attraction $V_{\text{eh}}$. Note that the above formalism should be understood as a functional integral: $Z=\int D[\psi,\Delta] e^{i\int dt dr L_e}$. Integrating out the fermions and keeping the quadratic terms in $\Delta$, one obtains the effective action:
\begin{align}
S[\Delta] =  \sum_{k_1, k_2, q} \text{Tr}
\left[  \hat{\Delta}_{k_1,q}^\dagger 
	V^{-1}_{k_1,k_2} \hat{\Delta}_{k_2,q} 
+
\sum_{\omega} \chi_{\Delta}(\omega,k,q) \hat{\Delta}^\dagger_{k,q}(\omega) \hat{\Delta}_{k,q}(\omega)
\right] ,
\quad
\chi_{\Delta}(\omega,k, q) = \frac{1}{-\omega+\xi_{k-q/2}+\xi_{k+q/2}}
\,.
\end{align}
Taking the saddle point equation $\frac{\delta S}{\delta \Delta_{k,q}^\dagger}=0$ and defining $\Delta_{k_,q}=\sum_{k^\prime} V_{k,k^\prime} \varphi_{k^\prime} \Phi_q$ where $\Phi_q$ means the excitonic field at momentum $q$ and $\varphi_{k^\prime}$ has the meaning of the wave function of the relative coordinate between the electron and the hole, one obtains the eigen mode equation
\begin{align}
\sum_{k^\prime} (2\xi_k \delta_{k k^\prime}+\frac{q^2}{4m}+V_{k,k^\prime}) \varphi_{k^\prime} =\omega \varphi_k
\,.
\label{eqn:bound_state}
\end{align}
It is the same as the Schrodinger equation for the two body bound state. Its eigen-modes $\varphi_{n}$ with eigen-values $\omega_{\text{exn}}+q^2/(4m)$ correspond to the excitons labeled by `$n$'.
Therefore, the excitonic eigen modes correspond to the two body bound states, and the order parameter field can be expanded using the eigen modes as 
\begin{align}
\Delta_{k_,q}=\sum_{n, k^\prime} V_{k,k^\prime} \varphi_{n, k^\prime} \Phi_{n,q}=\sum_n \left( \omega_{\text{exn}}-2\xi_k \right) \varphi_{n, k} \Phi_{n,q}
\end{align}
where $\Phi_{n,q}$ is the canonical bosonic field of the $n$th exciton  with center of mass momentum $q$, and $\omega_{\text{exn}}$ is its energy at zero $q$. 

In the following, we focus on the lowest $s$-exciton and drop the index `$n$', and add back the  interlayer tunneling $\hat{t}$, the electron-electron and hole-hole repulsion $V_{\text{ee}}, V_{\text{hh}}$ and  the electron-hole exchange $V_{\text{ex}}$.
The decomposed Lagrangian for coupled fermions and excitonic fields is simplified to 
\begin{align}
	L_{e} &= \psi^\dagger
	\begin{pmatrix}
		-i\partial_t +	\xi(p) & \hat{t}_p + \hat{\Delta}_p
		\\
		\hat{t}_p^\dagger + \hat{\Delta}^\dagger_p & -i\partial_t -\xi(p)
	\end{pmatrix}
	\psi
	+\sum_k (\omega_{\text{ex}}-2\xi_k) |\varphi_k|^2 \text{Tr}[\Phi^\dagger \Phi]
	+V_{\text{ee}} +V_{\text{hh}} + V_{\text{ex}} 
	\,, \notag \\
\hat{\Delta}_k &= (\omega_{\text{ex}}-2\xi_k) \varphi_k  \Phi
\end{align}
where we have suppressed the center of mass momentum $q$ for notational simplicity.
The excitonic action is obtained by integrating out the Fermions:
\begin{align}
	S =&  \sum_{q} c_2(q) \text{Tr} \left[ \Phi^\dagger(-q)   \Phi (q) 	\right] 
+
	\sum_{\sum q_i=0} c_4(q_1, q_2,q_3, q_4) \text{Tr}  \left[ \Phi^\dagger(q_1) \Phi(q_2) \Phi^\dagger(q_3) \Phi(q_4)
	\right]
+  \frac{1}{2} \int dr dr^\prime \rho(r) U(r-r^\prime) \rho(r^\prime)
\notag \\
	&+ \sum_{q} c_{\text{J}}(q)  
	\left[ \Phi_{12}(-q)\Phi_{21}(q) + c.c.+
	\sum_{ij} \Phi_{ij}^\ast(-q) \Phi_{ij}(q)  
	\right]
+  \sum_{q} c_{\text{em}}(q)  
\left[ 
	A_x (-q)  \partial_t \Phi_{0}(q) + iA_y(-q)   \partial_t \Phi_{3}(q) + c.c. 
\right]
	\notag\\
&+ \delta_{\text{ex0}} \left(\Phi_0^\ast \Phi_0 + \Phi_3^\ast \Phi_3 \right)
 \,
 \label{eqn:exciton_action_SI}
\end{align}
where $q$ should be understood as $(q, \omega)$.

\begin{figure}
	\includegraphics[width=0.8\linewidth]{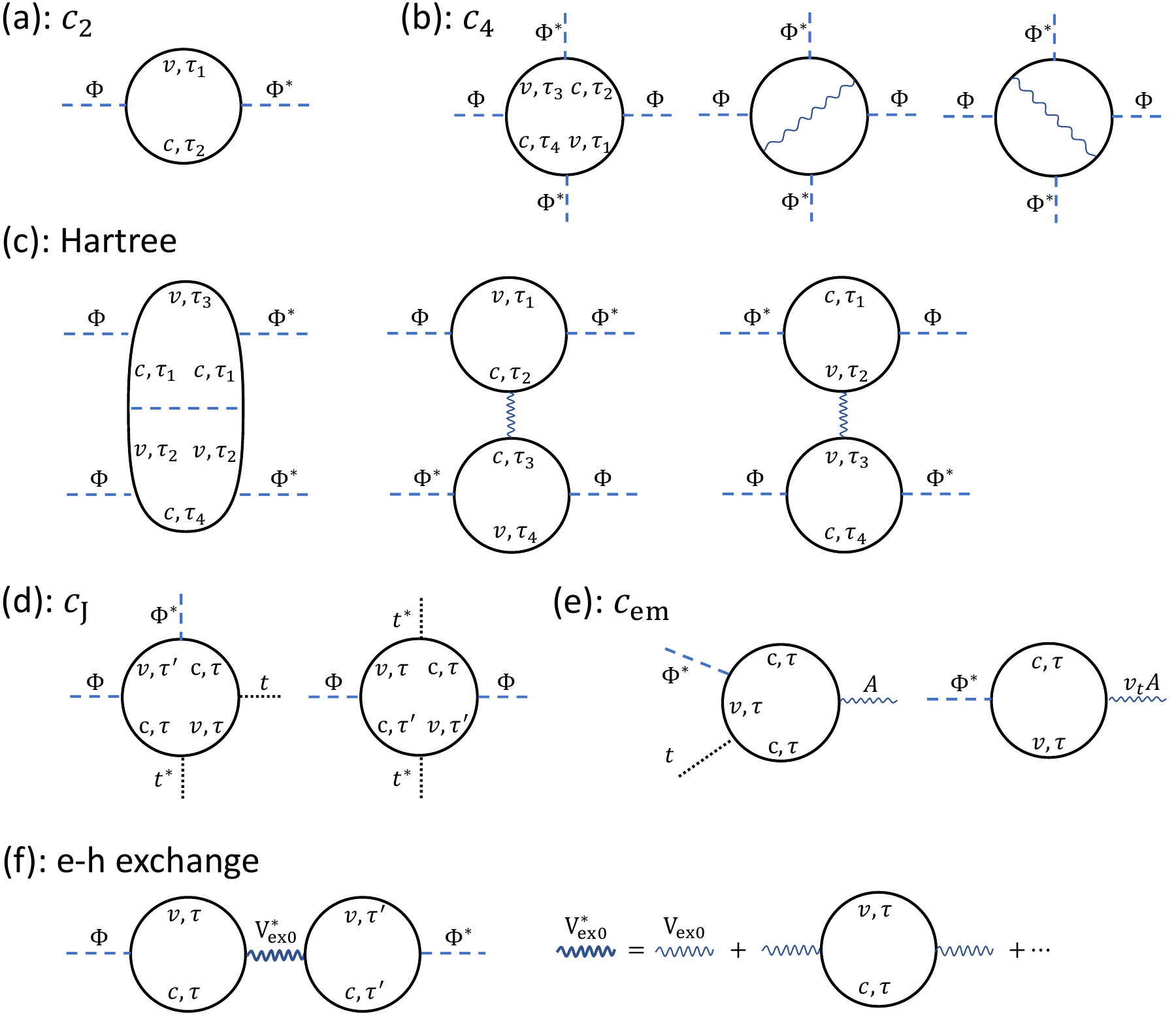} 
	\caption{The diagrammatic representation of each term in  \equa{eqn:exciton_action_SI}. Black solid lines are electron propagators. Blue dashed lines are order parameter fields. Black dashed lines are the interlayer tunneling $\hat{t}$. Wavy lines connecting two fermion lines are electron-electron or hole-hole interactions or electron-hole exchange interactions. Open wavy lines are EM vector potentials, meaning the current operator. The `Hartree' diagrams are for the direct dipole interaction between excitons (last term in \equa{eqn:exciton_action_SI}). Note that all of them are `one particle irreducible' with respect to the exciton field (blue dashed line). }
	\label{fig:Feynman_diagram}
\end{figure}

We now compute the coefficient of each term of the excitonic action \equa{eqn:exciton_action_SI},  diagrammatically represented in \fig{fig:Feynman_diagram}. 
In the simplest case, we neglect the screening from higher energy excitations which changes the interaction kernel to the Keldysh potential \cite{Keldysh1979}, and assume that the interlayer distance is much smaller than the exciton size, such that  the electron and hole attracts via the  two dimensional Coulomb potential, $V_q=2\pi e^2/(\epsilon q)$, where $\epsilon$ is a constant dielectric screening.  
The s-exciton wave function satisfying the bound state condition \equa{eqn:bound_state} has the the analytical form 
\begin{align}
\varphi_k=\sqrt{8\pi} \frac{q_0^2}{(q_0^2+k^2)^{3/2}}
\end{align} 
in two dimensions where $q_0=2/a_0$, $a_0=2\epsilon \hbar^2/me^2$ is the exciton Bohr radius,  and $E_{\text{b}}=me^4/(\hbar^2 \epsilon^2)=\hbar^2 q_0^2/m$ is the binding energy \cite{Yang1991}.
In the following integrals, there are apparent divergence as $G\rightarrow 0$. This is not physical because when $G$ is too small or when it is negative, the system crossovers to the BCS regime even for moderate $\mu$, such that  the power expansion of the action in $\Phi$ fails. We do not address this regime explicitly.

The $c_2$ term is the second term of \equa{eqn:exciton_action_SI} plus the bubble diagram in \fig{fig:Feynman_diagram}(a):
\begin{align}
	c_2(\omega, q) &= \sum_k (\omega_{\text{ex}}-2\xi_k) |\varphi_k|^2 
    - \sum_k (\omega_{\text{ex}}-2\xi_k)^2 |\varphi_k|^2 \frac{1}{\omega-\xi_{k+q/2}-\xi_{k-q/2}}
\notag\\
	&= \left(\omega-(\omega_{\text{ex}}+\frac{q^2}{4m}) \right) 
	\sum_k |\varphi_k|^2
  \frac{\omega_{\text{ex}}-2\xi_k}{\omega-\frac{q^2}{4m}-2\xi_k} 
   \xrightarrow{\omega \approx \omega_{\text{ex}} +\frac{q^2}{4m}} \omega-\left(\omega_{\text{ex}}+\frac{q^2}{4m} \right)
 \,
\end{align}
which gives the first term in \equa{eqn:exciton_Lagrangian} of the main text.

The $c_4$ term contains the positive `interlayer' exchange (first diagram in \fig{fig:Feynman_diagram} (b)) and negative intralayer exchanges (second and third diagrams) \cite{ Ciuti1998,Combescot2015,Wu2015}. Note that regardless of the terminology, these exchanges don't require any overlap between the atomic orbitals between the two layers. Here we compute the interlayer exchange at the zero momentum limit only: 
\begin{align}
	c_4 &= \sum_k |\varphi_k|^4 (\omega_{\text{ex}}-2\xi_k)^4 \frac{2}{(-\omega+2\xi_k)^3}
= (2\pi)^4 \frac{8E_{\text{b}}^4}{\pi^2 m^2} \sum_k  \frac{1}{(E_{\text{b}}+2\varepsilon_k)^{2}} \frac{1}{(\omega_{\text{ex}}-\omega+E_{\text{b}}+2\varepsilon_k)^3}
\notag\\
&\xrightarrow{\omega \approx \omega_{\text{ex}}}
\frac{16 \pi^2 \nu }{ m^2} = \frac{8\pi}{ m} =  \frac{4}{ \nu} 
\,
\end{align}
where $\nu=m/2\pi$ is the electronic density of states of the conduction/valence band.
With the intralayer exchanges included, the $c_4$ term can be estimated from dimensional analysis. It can only be a function of $m$, $e$, $\hbar$ and the interlayer distance $d$, which means $c_4 \sim \frac{1}{m} f(\frac{d}{a_0})$ from dimensional analysis where $a_0=\hbar^2/(me^2)$. In the limit of small interlayer distances, $d$ drops out  such that  $c_4 \sim \frac{1}{m}$ which does not even depend on $e$. As shown in Refs.~\cite{Combescot2015,Wu2015}, as $d/a_0$ increases, $f(\frac{d}{a_0})$ crossovers from $\sim 1$ to negative values.
Therefore, for adjacent bilayers such that the interlayer distance is small, the positive interlayer exchange is generally larger in magnitude than the intralayer exchange, rendering the net exchange interaction a repulsive one \cite{Wu2015}.

The  Hartree diagrams  (\fig{fig:Feynman_diagram}(c)) simply give the long range dipole-dipole interaction $U(r) \approx \frac{d^2}{r^3}$ between the excitons due to their permanent interlayer dipoles.

The $c_{\text{J}}$ term is the second order Josephson  term represented by \fig{fig:Feynman_diagram}(d) where the first diagram gives $c_{\text{J}} v_{\text{t}}^2 \rho$ and the second diagram gives $c_{\text{J}} v_{\text{t}}^2 (\Phi_{12}\Phi_{21}+c.c.)$ since the intravalley terms vanish due to the chiral momentum dependence of $t_{\text{k}}$.  The coefficient is 
\begin{align}
c_{\text{J}}(\omega) &= v_{\text{t}}^2 \sum_k |\varphi_k|^2 k^2  \frac{(\omega_{\text{ex}}-2\xi_k)^2}{\xi_k (4\xi_k^2-\omega^2)}
 = 
4\pi^2 v_{\text{t}}^2 \sum_k \frac{4}{\pi} E_{\text{b}}^2
\left\{
\begin{array}{lc}
  \frac{ 2\varepsilon_k}{(E_{\text{b}}+2\varepsilon_k)^2 (G+2\varepsilon_k) (2G-E_{\text{b}}+2\varepsilon_k)}  \,,
	& \omega = \omega_{\text{ex}} \\
\frac{2\varepsilon_k}
{(E_{\text{b}}+2\varepsilon_k) (G+2\varepsilon_k)^3} \,,
	&  \omega =0 
\end{array}
\right.
\notag\\
& = 
4\pi^2 \frac{2}{\pi} v_{\text{t}}^2 \nu E_{\text{b}}^2
\left\{
\begin{array}{lc}
 \frac{
	2(G-E_{\text{b}}) + E_{\text{b}} \ln \frac{E_{\text{b}}}{2G-E_{\text{b}}} + 2G \ln \frac{E_{\text{b}}(2G-E_{\text{b}})}{G^2}
}
{4(E_{\text{b}}-G)^3}  \,,
	& \omega = \omega_{\text{ex}} \\
 \frac{
	E_{\text{b}}^2 - G^2 + 2 G E_{\text{b}} \ln \frac{G}{E_{\text{b}}} 
}
{2(E_{\text{b}}-G)^3G} \,,
	&  \omega =0 
\end{array}
\right.
\notag\\
&=m v_{\text{t}}^2  f_{\text{J}}(E_{\text{b}}, G, \omega)
\,,
\notag\\
f_{\text{J}}
&\rightarrow
4\pi^2 \frac{ E_{\text{b}}^2}{G^2} 
	\left\{
	\begin{array}{lc}
		\frac{1}{2\pi^2} \ln \frac{G}{E_{\text{b}}} \,,
		& G\gg E_{\text{b}},\, \omega \approx \omega_{\text{ex}} \\
		\frac{1}{6\pi^2} \,,
		& G\approx E_{\text{b}} ,\, \omega \approx 0  \\
		\frac{1}{2\pi^2} \,,
        & G\gg E_{\text{b}} ,\, \omega \approx 0 
	\end{array}
	\right..
\label{eqn:c_J_appendix}
\end{align}
Note that the $\Phi$ legs  \fig{fig:Feynman_diagram}(d) have frequencies $\pm \omega$.

The $c_{\text{em}}$ term  (\fig{fig:Feynman_diagram}(e))  gives the coupling to the in-plane electric field represented by the dynamical vector potential $(A_x,\,A_y)$. The physical meaning is that dynamics of the excitonic field is accompanied by in plane electrical currents, i.e., the s-excitons are optically active.  The sum of the `intraband current' (left diagram of \fig{fig:Feynman_diagram}(e)) and `interband current' (right diagram of \fig{fig:Feynman_diagram}(e)) gives the coefficient
\begin{align}
	c_{\text{em}}(\omega) &= \frac{1}{2} v_{\text{t}} \sum_k \varphi_k  \frac{(\omega_{\text{ex}}-2\xi_k)(v_xk_x-\xi_k)}{ (\omega-2\xi_k)\xi_k^2} 
= 2\pi v_{\text{t}} 
    \frac{1}{2} \sum_k \sqrt{\frac{2}{\pi}} \frac{q_0^2}{(q_0^2+k^2)^{3/2}} \frac{(E_{\text{b}}+2\varepsilon_k)(G/2)}{ (\omega-2\xi_k)\xi_k^2} 
\notag\\
     &= -2\pi v_{\text{t}}  \frac{GE_{\text{b}}}{\sqrt{m}} \sqrt{\frac{2}{\pi}} \sum_k  \frac{1}{(E_{\text{b}}+2\varepsilon_k)^{1/2} (-\omega+G+2\varepsilon_k)(G+2\varepsilon_k)^2} 
 \notag\\
 &=\sqrt{m v_{\text{t}}^2/E_{\text{b}} }
 f_{\text{em}}(E_{\text{b}}, G, \omega)
 \,, \notag\\
 f_{\text{em}} &\rightarrow 2\pi 
\frac{GE_{\text{b}}^{3/2}}{(2\pi)^{3/2}} 
 \left\{
 \begin{array}{lc}
 \frac{(E_{\text{b}}+2G)(E_{\text{b}}-G)+3G\sqrt{E_{\text{b}}(G-E_{\text{b}})} \arccos(\sqrt{E_{\text{b}}/G})}{\sqrt{E_{\text{b}}} G (E_{\text{b}}-G)^3}  \,,
 &  \omega_1 \approx \omega_{\text{ex}} \\
 \frac{2E_{\text{b}}^{3/2}\sqrt{G-E_{\text{b}}}-5G\sqrt{E_{\text{b}}(G-E_{\text{b}})}+3G^2 \arccos(\sqrt{E_{\text{b}}/G})}{4G^2 (-E_{\text{b}}+G)^{5/2}}  \,,
 	& \omega \approx  0  
 \end{array}
 \right.
 \notag\\
 &\rightarrow 
 \frac{2  }{(2\pi)^{1/2} } 
\frac{E_{\text{b}}}{G} 
 \left\{
\begin{array}{lc}
	 1  \,,
	& G\gg E_{\text{b}},\, \omega \approx \omega_{\text{ex}} \\
1/5 \sqrt{\frac{E_{\text{b}}}{G}}  \,,
	& G\approx E_{\text{b}} ,\, \omega \approx  0  \\
	\frac{3\pi}{16} \sqrt{\frac{E_{\text{b}}}{G}}   \,,
	& G\gg E_{\text{b}} ,\, \omega \approx  0 
\end{array}
	\right..
\label{eqn:cem_appendix}
\end{align}
Note that to derive the above expression, we subtracted  the $\omega=0$ limit of the diagrams to obtain the current proportional to $\omega$, which gives the $\partial_t$ in \equa{eqn:exciton_action_SI}. This `renormalization' works because the current response to a static field should be zero.

The $\delta_{\text{ex0}}$ term comes from the weak electron-hole exchange $V_{\text{ex}}$ ( \fig{fig:Feynman_diagram}(f)) that requires a nonzero overlap between the conduction and valence band orbitals residing on different layers:
\begin{align}
	\delta_{\text{ex0}} (\omega) &= V_{\text{ex0}}^\ast  (\omega)
	\left[\sum_k  \varphi_k \frac{\omega_{\text{ex}}-2\xi_k}{\omega-2\xi_k}
		\right]^2
	\xrightarrow{\omega \rightarrow \omega_{\text{ex}}} V_{\text{ex0}}^\ast 
	\left[\sum_k  \varphi_k
	\right]^2
	= 128 \pi^3 V_{\text{ex0}}^\ast \frac{1}{a_0^2}
,\notag \\
V_{\text{ex0}}^\ast  (\omega)	&= \frac{V_{\text{ex0}}}{1+V_{\text{ex0}} 2\sum_k \frac{1}{\omega-2\xi_k}}
\approx  V_{\text{ex0}}
	\,.
\end{align}

 Taking the appropriate frequency arguments and neglecting the $\delta_{\text{ex0}}$  term in \equa{eqn:exciton_action_SI}, one obtains \equa{eqn:exciton_Lagrangian} in the main text. Note that the above formalism works for both interlayer excitons in bilayers and intralayer excitons in monolayers. In the latter case, one may choose to work in the band diagonalized basis where there is no explicit interband tunneling term for the electrons, but the interaction will have interband terms that contribute the same Josephson couplings in \equa{eqn:exciton_Lagrangian} of the main text.

\section{Dynamical orbits and the effective potential}
\begin{figure}
	\includegraphics[width=0.85\linewidth]{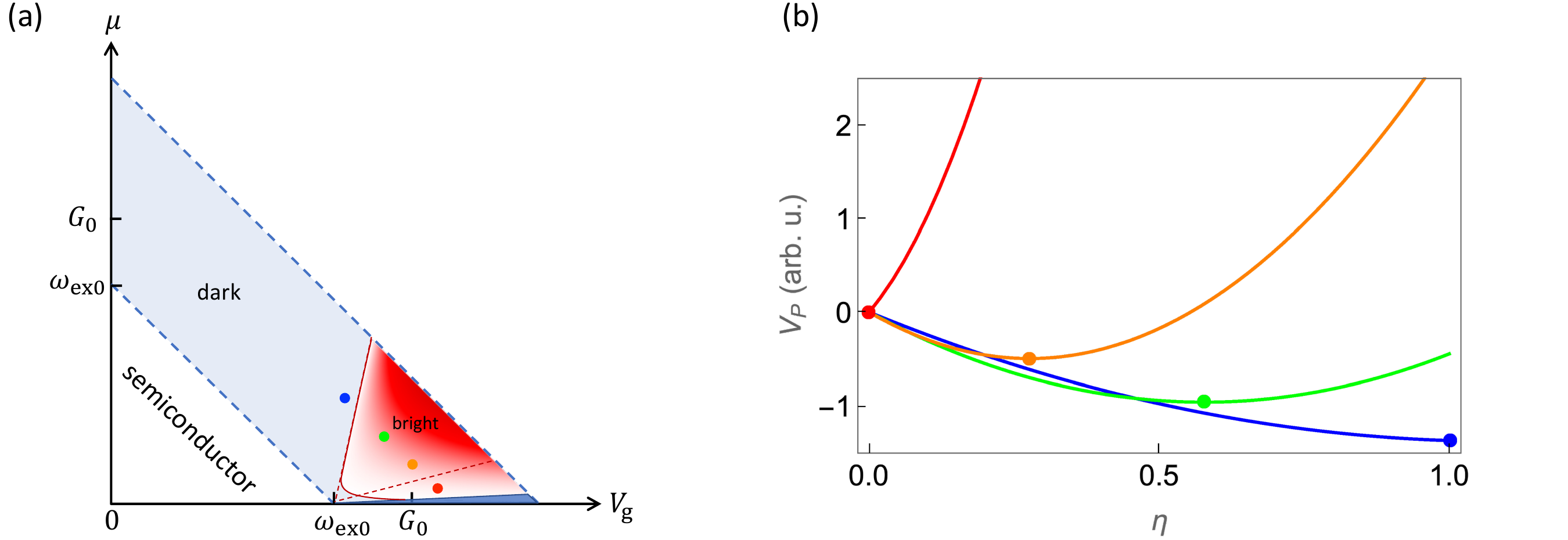} 
	\caption{(a) Four representative points are chosen on the non-equilibrium phase diagram. (b) The potential $V_{\text{P}}(\eta)$ plotted for the four points in (a) with correspondence labeled by their colors. The dots in (b) are the minima of the curves.
	}
	\label{fig:vp_eta}
\end{figure}

In this section, we solve the equation of motion
$
	i\partial_t \Phi = \partial_{\Phi^\dagger} H -  i\Gamma \partial_{\Phi^\dagger} F  
$
to obtain the leading order correction of $H_J$ to the dynamical orbit of the order parameter, and to compute the effective potential that determines the phase diagram.
Changing the variables into phases and amplitudes: $(\Phi_1, \Phi_3)=  (\sqrt{\eta}, \sqrt{1-\eta}e^{i\theta_a}) \sqrt{\rho} e^{i\theta}$ (\equa{eqn:eta} of the main text), the Lagrangian density from \equa{eqn:exciton_Lagrangian} is written as
\begin{align}
L= \rho \dot{\theta} + (1-\eta) \rho  \dot{\theta}_a
+ \omega_0 \rho + g\rho^2
+c_4 \rho^2 \eta (1-\eta) \sin^2 \theta_a
+ c_{\text{J}} \eta \rho \cos 2\theta
\,.
\label{eqn:L_theta_eta}
\end{align} 
where $g=(U_0+c_4)/2$.
The equation of motion \equa{eqn:EOM_eta} is expanded as 
\begin{align}
&\dot{\theta} =- \omega_0-2g\rho 
 -(1-\eta)  \dot{\theta}_a
-2 c_4 \rho \eta (1-\eta) \sin^2 \theta_a 
-  c_{\text{J}} \eta  \cos 2\theta
,\quad
\dot{\rho} = -2 c_{\text{J}} \eta \rho \sin 2\theta
+ 4 \rho_0\Gamma (-\partial_\rho H+\mu) ,\quad
\notag\\ 
& \rho \dot{\theta}_a =
c_4 \rho (1-2\eta) \sin^2 \theta_a 
+c_{\text{J}} \rho  \cos 2\theta
,\quad
\partial_t\left[ \rho (1-\eta)  \right] = 
c_4 \rho^2 \eta (1-\eta) 2 \sin \theta_a \cos \theta_a
	\,.
	\label{eqn:EMO_eta}
\end{align} 
Note that the main effect of the device-bath tunneling $\Gamma$ is to maintain the chemical potential $\mu$. As long as it exists and is not too large, it does not affect the phase diagram qualitatively. 
For simplicity, we keep only the  amplitude ($\rho$) direction of the damping term $\Gamma \partial_{\Phi^\dagger} F \rightarrow \Gamma \partial_{\Phi^\dagger} F|_\rho$, which does not change the qualitative conclusion. 

The solution (orbit) to \equa{eqn:EMO_eta} may be written as  $S=\left(\theta, \rho, \theta_a, \eta \right)_t=
S_0+
\left(\theta^{(1)}, \rho^{(1)}, {\theta}^{(1)}_a, \eta^{(1)} \right)
+ O(c_{\text{J}}^2)
$, where $S_0=\left(\theta_0(t), \rho_0, 0, \eta_0 \right)$ is the zeroth order orbit and $\theta_0(t)=-\mu t$. The $O(c_{\text{J}})$ correction is an oscillating term with frequency $\pm 2\mu$. Its $2\mu$ components are 
\begin{align}
&{\theta}^{(1)}_a(2\mu) =\frac{i}{4\mu} c_{\text{J}}
,\quad
{\rho}^{(1)}(2\mu)  =  \frac{\eta}{-i2\mu+ \gamma} 
\left( 
- \frac{\gamma}{2g} 
+ 2i  \rho
\right) \frac{c_{\text{J}} }{2}
,\quad
\notag\\ 
&{\theta}^{(1)}(2\mu) =
\frac{i}{2\mu} 
\left( 
-2g {\rho}^{(1)}(2\mu) -   \frac{c_{\text{J}} }{2} 
\right)
,\quad
{\eta}^{(1)}(2\mu) = (1-\eta)\frac{{\rho}^{(1)}(2\mu)}{\rho_0}
+\frac{1}{i2\mu} c_4 \rho \eta (1-\eta) 2 {\theta}^{(1)}_a(2\mu)
\,
	\label{eqn:orbit_cj}
\end{align} 
where we have defined the damping rate $\gamma=8g \rho_0 \Gamma$.
The static (Ponderomotive) force for $\eta$ is thus
\begin{align}
F_{\text{P}}(\eta) &= -\langle \partial_\eta L\rangle_t 
=  
	-\langle 
	c_{\text{J}}  \rho \cos 2\theta - \rho \dot{\theta}_a
	+ c_4  \rho^2 (1-2\eta) \sin^2 \theta_a
	\rangle_t 
	\notag\\
& 	\xrightarrow{O(c_{\text{J}}^2)}
-c_{\text{J}} \frac{1}{2} \left(
{\rho}^{(1)}(2\mu) -i 2\rho {\theta}^{(1)}(2\mu) + c.c.
\right)
+\left(
{\rho}^{(1)}(2\mu) i 2\mu \theta_a(-2\mu)
+c.c.
\right)
-c_4 \rho_0^2 (1-2\eta) \frac{c_{\text{J}}^2}{8\mu^2}
\notag\\
&=  
-c_{\text{J}} \frac{1}{2} \left(
-i 2\rho_0 {\theta}^{(1)}(2\mu) + c.c.
\right)
-c_4 \rho_0^2 (1-2\eta) \frac{c_{\text{J}}^2}{8\mu^2}
=  
c_{\text{J}}^2  \rho_0
\left[ 
\frac{ 2\mu-4g \rho_0 \eta  + (1-\eta)\frac{\gamma^2}{2\mu} }{4\mu^2 + \gamma^2}
+ c_4 \rho_0 \frac{2\eta-1}{8\mu^2} 
\right]
\,
\label{eqn:p_force_SI}
\end{align}
which is just \equa{eqn:p_force} of the main text.
Note that naively speaking, $ \partial_\eta L  $ should be zero at every order of $c_{\text{J}}$ since this is what the equation of motion (the third equation in \equa{eqn:EMO_eta}) says. From this perspective, the static force $F_{\text{P}}$ should act on the variable $\theta_a$ at order  $c_{\text{J}}^2$, giving $\dot{\theta}_a = F_{\text{P}}$. However, note that there is a strong potential term $c_4 \rho^2 \eta (1-\eta) \sin^2 \theta_a$ in \equa{eqn:L_theta_eta} that bounds $\theta_a$, meaning that the `zero frequency' component of $\theta_a$ cannot be increasing.
Moreover, there are actually dissipative terms in the equation of motion for $\theta_a$ and $\eta$ due to the environment: $\dot{\theta}_a -\Gamma \dot{\eta} \sim F_{\text{P}}/\rho_0$, $-\dot{\eta} + \Gamma \dot{\theta_a} \sim  \theta_a$. As a result, the slow component of $\theta_a$ stabilizes at a certain value $\sim F_{\text{P}}/\Gamma$, and the static force $F_{\text{P}}(\eta)$ equals $-\Gamma \partial_t \eta$. Therefore, $F_{\text{P}}$ is the actual Ponderomotive force for $\eta$ .

\emph{The phase diagram---}The Pondermotive force corresponds to an effective potential for $\eta$, coined the Pondermotive potential~\cite{Sun.2023_ponderomotive}: $V_{\text{P}}(\eta) = - \int d\eta  F_{\text{P}}$. Its value of the bright state relative to the dark one is found to be 
\begin{align}
	\delta V_{\text{P}} =V_{\text{P}}(0) - V_{\text{P}}(1) &=  c_{\text{J}}^2   \rho_0 \frac{\mu+\omega_{\text{ex}}+\frac{\gamma^2}{4\mu}}{4\mu^2 +\gamma^2} 
	\,.
	\label{eqn:p_potential}
\end{align}
In the limit of $\gamma=0$, $\delta V_{\text{P}}$ equals the time-averaged Lagrangian~\cite{Sun.2023_ponderomotive} of the bright state relative to the dark one, see \equa{eqn:mean_field_driving_energy_1}.
If one further takes $\mu = \omega_{\text{ex}}$, it reduces to the energy splitting between a single bright and dark exciton due to the $c_{\text{J}}$ term.

Minimizing $V_{\text{P}}(\eta)$ gives the mean field steady state, yielding the phase diagram. Surprisingly, in most of the `bright' phase of \fig{fig:vp_eta}(a),  the energy minimum lies somewhere at ($0<\eta_0<1$)  determined by $F_{\text{P}}(\eta_0)=0$, as shown by the profile of the potential in \fig{fig:vp_eta}(b). This means that the order parameter is a coherent hybrid of the bight and dark condensate. Since this mixture emits light, we still call it a `bright' condensate although it also has a dark component.  

In the limit of $\gamma \rightarrow 0$, the phase diagram is simpler to discuss. The region of the hybrid bright state is $- \frac{\frac{c_4}{2g}}{4-\frac{c_4}{2g}} \omega_{\text{ex}}<\mu< -(\frac{8g}{c_4}-1) \omega_{\text{ex}}$ whose boundary is shown by the two red dashed lines in \fig{fig:vp_eta}(a). Here one has $F_{\text{P}}(0)>0$ and $F_{\text{P}}(1)<0$, so that the energy minimum lies at 
\begin{align}
\eta_0= \frac{\frac{\mu}{\mu-\omega_{\text{ex}}}-\frac{c_4}{8g}}{1-\frac{c_4}{4g}}
= \frac{\frac{\mu}{\mu-\omega_{\text{ex}}}-\frac{c_4}{4(c_4+U_0)}}{1-\frac{c_4}{2(c_4+U_0)}}
\xrightarrow{U_0 \ll c_4}
\frac{2\mu}{\mu-\omega_{\text{ex}}}-\frac{1}{2}
	\,,
\label{eqn:eta0}
\end{align}
as shown in \fig{fig:vp_eta}.
At the upper and lower boundaries, the phase transitions are continuous, and the hybrid bright condensate breaks more symmetries than the `pure' condensates.
In the experimentally relevant limit of $U_0 \ll c_4$, the region of the hybrid bright state simplifies to $- \frac{1}{3} \omega_{\text{ex}}<\mu< -3 \omega_{\text{ex}}$. The condensate is a pure bright one in the lower region $\mu< - \frac{1}{3} \omega_{\text{ex}}$ and a pure dark on in the upper region $\mu> - 3 \omega_{\text{ex}}$.

Note that for a constant dimensionless $\Gamma$ in Eq.~2 of the main text, the physical tunneling rate $\gamma=8g \rho_0 \Gamma$ defined below Eq.~10 of the main text  should depend on the exciton density. However, this only weakly affects the phase diagrams in Fig.~3(a) of the main text and \fig{fig:vp_eta}(a) (bottom left corner of the `bright' region), and we plot
the phase diagrams   for a constant $\gamma=0.1 \unit{eV}$ for simplicity. 

\subsection{The effect of  light emission}
In this subsection, we discuss the radiation of free space photons due to the $c_{\text{em}}$ term in the bright condensate. We show that because $c_{\text{em}}$ is suppressed by the fine structure constant $\alpha$, its effect on the Ponderomotive potential is small compared to the $c_{\text{J}}$ term, such that its effect on the phase diagram can be neglected for the device in free space.

The phase winding of the bright condensate leads to oscillation of the in-plane electrical polarization 
$(P_x, P_y)= 2 c_{\text{em}}  (-\text{Re}[\Phi_0], \text{Im}[\Phi_3])$
which emits free space radiation with the electric field $E=\frac{2\pi}{c} \partial_t P$ (a result of solving the Maxwell's equation with the correct boundary conditions on the 2D plane), see  \fig{fig:cavity}(a). Therefore, the electric field has a $\pi/2$ phase difference with the polarization. In other words, it is anti-parallel with the oscillating current, doing negative work to the system whose power is just the radiation power:
\begin{align}
	P_{\text{r}}&=\frac{c}{4\pi}E^2 = \gamma_r \mu \rho
\,,\quad
\gamma_r = \frac{4\pi}{c} c_{\text{em}}^2  \mu
\approx 2 \alpha \frac{m v_{\text{t}} ^2 E_{\text{b}}}{G^2} \mu
	\label{eqnSI:radiation_power}
\end{align}
which also implies a tunneling charge current: $I_{\text{bright}}=P_{\text{r}}/\mu=\gamma_r \rho$.  
The electric field feeds back on the dynamics through the $c_{\text{em}}$ term in \equa{eqn:exciton_Lagrangian} of the main text, and distorts the orbit.  
Note that the magnetic field of the emitted light is in-plane and does not couple to the $\Phi_0$ or $\Phi_3$ excitons, since their in-plane spin magnetic moment is zero.

To compute the distorted orbit, we consider the pure bright condensate for simplicity with, e.g., only $\Phi_3=\sqrt{\rho} e^{i\theta}$ nonzero. Similar to the dark condensate, the orbit wiggles around  the perfect circle $S_0=\left(\theta_0(t), \rho_0 \right)$ due to the feed back force of the emitted electric field. However, the main effect is an overall reduction of the exciton density due to the photon emission. For simplicity, we assume an  incoherent emission that provides a constant friction to reduce the density with rate $\gamma_r$, and the equation of motion is simplified to
\begin{align}
	&\dot{\theta} =- \omega_0-2g\rho 
	,\quad
	\dot{\rho} = -\gamma (\rho-\rho_0) -\gamma_r \rho 
\,.
\label{eqn:EMO_bright}
\end{align} 
The solution is a smaller orbit as shown in \fig{fig:Static_condensate}(a) of the main text, with the density reduced by $\delta \rho \approx \frac{\gamma_r}{\gamma} \rho_0$. Its correction to the average free energy is $\delta F_{\text{EM}} \approx g \delta \rho^{ 2} \sim  g\rho_0^2 \alpha^2   \frac{m^2 v_{\text{t}}^4 E_{\text{b}}^2}{G^4} \frac{\mu^2}{\gamma^2}$. Compared to the effective potential ($\delta V_{\text{P}}$ in \equa{eqn:p_potential}) caused by the Josephson coupling  $c_{\text{J}}$, it is smaller by a factor of 
\begin{align}
\delta F_{\text{EM}}/ \delta V_{\text{P}} 
\xrightarrow{\gamma_R \ll \gamma \ll \mu}
 \frac{\mu^4}{E_{\text{b}}^2 \gamma^2}
\alpha^2
\xrightarrow{E_{\text{b}} \sim \mu}
 \frac{\mu^2}{\gamma^2}
\alpha^2
\,.
\label{eqn:cem_vs_cj}
\end{align} 
In the interested parameter regime, $\mu/\gamma \sim 10$ while the fine structure constant is $\alpha \approx 1/137$, leading to $\delta F_{\text{EM}}/ \delta V_{\text{P}}  \ll 1$. Although $\delta F_{\text{EM}}$ is not exactly the effective potential contributed by the $c_{\text{em}}$ term that the steady state minimizes, the latter should be at the same order. Therefore,  it is safe to neglect the $c_{\text{em}}$ terms when determining the phase diagram in \fig{fig:Dynamic_condensate}(a) of the main text.

\section{The device in a Fabry–Pérot cavity}
\begin{figure}
	\includegraphics[width=0.8\linewidth]{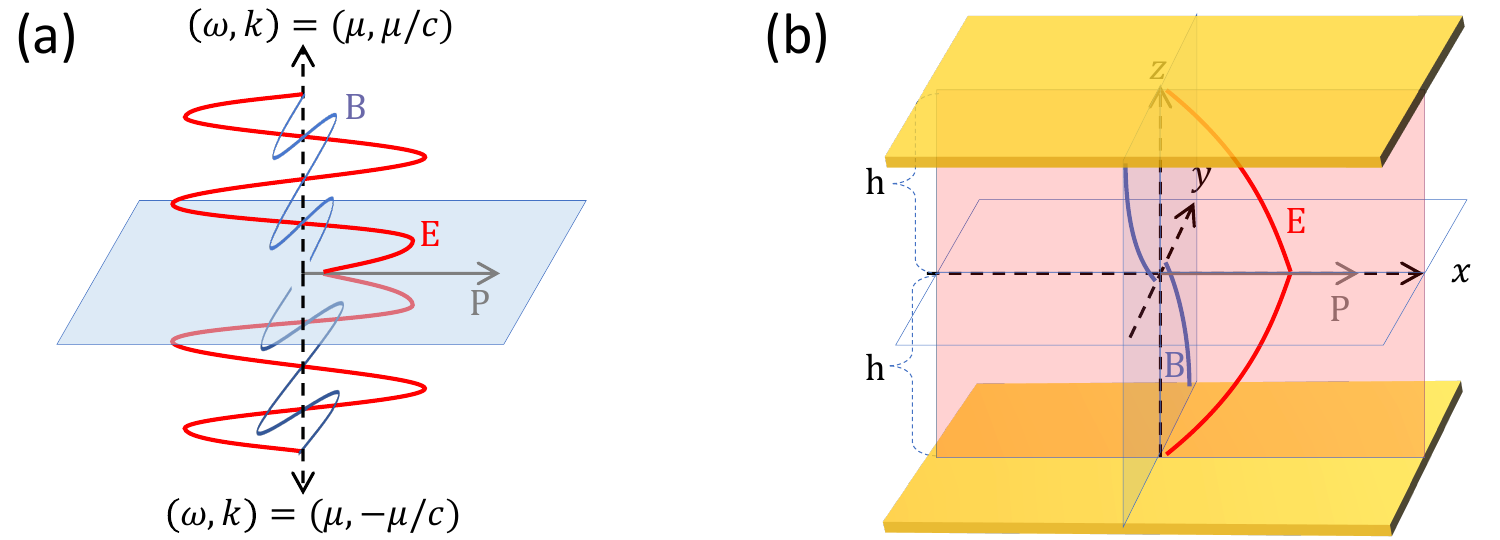} 
	\caption{(a) Illustration of the photon radiation of the dynamical bright condensate in free space.
		(b) Illustration of the 2D system (x-y plane) placed in the middle  of a  Fabry–Pérot cavity. The red/blue lines are the dynamical  electric/magnetic fields  in response to the dynamical 2D electrical polarization (gray arrow).
	}
	\label{fig:cavity}
\end{figure}

In this section, we compute the oscillating EM field in the cavity (\fig{fig:Dynamic_condensate}(b) in the main text and \fig{fig:Devices}) in response to the dynamical polarization $Pe^{-i\omega t}$ of the device, where $\omega=\mu$. This is done by solving the Maxwell's equation with appropriate boundary conditions. 

Without loss of generality, we assume the polarization is along the $x$ direction, which  induces an electric/magnetic field along the $x$/$y$ direction, as shown in \fig{fig:cavity}(b). Close to the mirrors ($z=-h$), the electric field vanishes, meaning $E_x(\omega)=E_0 \sin k(z+h)$ for $-h<z<0$ where $k=\sqrt{\epsilon} \omega/c$. This together with Maxwell's equation leads to the magnetic field $B_y(\omega)=-i\sqrt{\epsilon} E_0 \cos k(z+h)$. 

Right above and below the 2D plane, $E_x$ is continuous, so that the electric/magnetic field in the upper half space $0<z<h$ is symmetric/anti-symmetric with respect to that in the lower half space, as shown in \fig{fig:cavity}(b). Ampère's circuital law applied to the 2D plane leads to $-2i\sqrt{\epsilon} E_0 \cos kh=\frac{4\pi}{c}\partial_t P$, so that $E_0=\frac{4\pi \omega}{2c\sqrt{\epsilon}} P/\cos kh$. Therefore, the electric field on the 2D plane is $E=\frac{2\pi \omega}{\sqrt{\epsilon}c} P \tan kh = \frac{2\pi \omega}{\sqrt{\epsilon}c} P \tan \frac{\sqrt{\epsilon} \omega}{ch}$. To qualitatively incorporate the damping of the cavity modes, we simply replace $\omega$ by $\omega+i\gamma_{\text{p}}$.

An equivalent description is that, the cavity has  discrete photon modes which are Harmonic oscillators. Each oscillator is linearly driven by the dynamical polarization $Pe^{-i\omega t}$ and responds by contributing an EM field. The total response field is the sum of all of them, which is just the solution of Maxwell's equation.

\section{The rotating frame}
To compute the tunneling current of the dynamical condensate, it is more convenient to perform a gauge transformation in time to the `rotating' frame: $\Phi \rightarrow \Phi e^{-i 2\mu t}$. Afterwards, the exciton energy is shifted to $\omega_{\text{ex}} \rightarrow \omega_{\text{ex}}^\prime=  \omega_{\text{ex}}-\mu$ and $H_0$ in \equa{eqn:exciton_Lagrangian} leads to a static condensate if the parameters are in the colored the region of \fig{fig:Dynamic_condensate}, while $H_{\text{J}}$ becomes a periodically driving term:
\begin{align}
	H_0 = &  \text{Tr} \left[ \Phi^\dagger \left(\omega_{\text{ex}}^\prime  + \frac{p^2}{4m}
	\right)  \Phi + c_4 (\Phi^\dagger \Phi)^2
	\right]
	+ \frac{1}{2} \int dr dr^\prime \rho(r) U(r-r^\prime) \rho(r^\prime)
	\,,
	\notag\\
	H_{\text{J}} = &
	c_{\text{em}} \left( E_x  \Phi_{0} e^{-i\mu t} + iE_y  \Phi_{3}  e^{-i\mu t}  + c.c. \right) +
	c_{\text{J}} 
	\left[
	\frac{1}{2}(\Phi_1^2 + \Phi_2^2) e^{-i2\mu t} +c.c. +\rho
	\right] 
	\,.
	\label{eqn:H_drive}
\end{align} 
The equation of motion simplifies to 
\begin{align}
	i\partial_t \Phi = (1-  i\Gamma) \partial_{\Phi^\dagger} H  \,, \quad 
	H=	H_0  + 	H_{\text{J}}
	\label{eqn:EOM_SI}
\end{align}
where $H_0$ and $H_{\text{J}}$ are from \equa{eqn:H_drive}.
The system has now become formally a periodically driven system. Discussions in the following sections will be in the rotating frame.

In the condensed phase ($\omega_{\text{ex}}^\prime <0$), the mean field exciton density is $\rho_0=\Phi_\mu^\ast \Phi^\mu=-\omega_{\text{ex}}/(2g)$. Assuming a dark condensate with only $\Phi_1=\sqrt{\rho_0}$ nonzero, one may write the order parameter as its mean field value plus a small fluctuation: $\Phi_\mu+\delta \Phi_{\mu}$.  The equation of motion for $\delta \Phi_1$ at linear order is expanded from \equa{eqn:EOM_SI} as
\begin{align}
G^{-1}
\begin{pmatrix}
	\delta \Phi_{1}
	\\
	\delta \Phi_{1}^{\ast}
\end{pmatrix}	
=
\begin{pmatrix}
\frac{-i}{1-i\Gamma} \partial_t +  \frac{q^2}{4m} 
+ 2 g\rho_0 &  2 g\rho_0 
\\
2 g\rho_0 
& 
\frac{i}{1+i\Gamma} \partial_t +  \frac{q^2}{4m} 
+ 2 g\rho_0
\end{pmatrix}	
\begin{pmatrix}
	\delta \Phi_{1}
\\
	\delta \Phi_{1}^{\ast}
\end{pmatrix}	
=-	c_{\text{J}}  \sqrt{\rho_0}
\begin{pmatrix}
e^{i2\mu t} 
	\\
e^{-i2\mu t} 
\end{pmatrix}	
\end{align}
where the retarded Green's function in the frequency basis reads
\begin{align}
G(q,\omega)=
\frac{1}{
-\frac{\omega^2}{1+\Gamma^2}
+ (\frac{q^2}{4m} 
+ 2 g\rho_0 ) \frac{-2 i\Gamma \omega}{1+\Gamma^2}
+ \omega_q^2
}
\begin{pmatrix}
\frac{\omega}{1+i\Gamma}  +  \frac{q^2}{4m} 
+ 2 g\rho_0  & 
-2 g\rho_0  
\\
-2 g\rho_0  
		& 
\frac{-\omega}{1-i\Gamma}  +  \frac{q^2}{4m} 
+ 2 g\rho_0 
\end{pmatrix}	
\,
\label{eqn:green_function}
\end{align}
and 
\begin{align}
\omega_q= \sqrt{\frac{q^2}{4m} \left( \frac{q^2}{4m}+ 4g \rho_0 \right)}
\end{align}
is the phase mode frequency at momentum $q$ in the limit of $\Gamma=0$ and $H_{\text{J}}=0$.

\section{The tunneling current}
\begin{figure}
	\includegraphics[width=0.9 \linewidth]{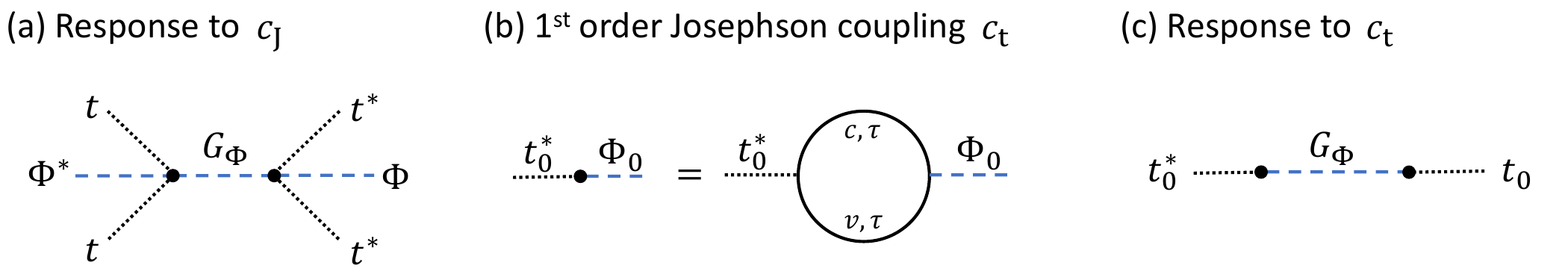} 
	\caption{(a) Linear response diagram of $\Phi$ to $t^2 e^{-i2\mu t}$ at the mean field level. The open $\Phi$ and $\Phi^\ast$ legs are  their mean field values. (b) The linear coupling  between the tunneling $t_0$ and the excitonic field $\Phi_0$ in the case of first order Josephson tunneling. (c) Linear response diagram of $\Phi$ to $t_0 e^{-i\mu t}$ at the mean field level. }
	\label{fig:driving_diagram}
\end{figure}

Since we focus on the dark condensate in this section, we  neglect the $c_{\text{em}}$ terms.  The driving term $H_{\text{J}}$ in \equa{eqn:H_drive} would inevitably do work to the system, whose power supply comes from the bath, i.e., the leads in the bilayer case. Therefore, the power dissipation of the driving term corresponds directly to a DC tunneling current. The simplest contribution comes from the mean field driving term in \fig{fig:driving_diagram}(a), which says $\Phi_1^2 e^{-2i\mu t} \rightarrow 2 \Phi_1  e^{-2i\mu t} \delta \Phi_1 $ and its complex conjugate linearly drives the zero momentum condensate. It gives	
\begin{align}
I_{\text{dark}}=\frac{1}{\mu}P_{\text{drive}} = \frac{1}{\mu} 2\mu c_{\text{J}}^2 \Phi^{\ast} \Phi \mathrm{Im}[G_\Phi(-2\mu)]
=  2 c_{\text{J}} \rho_0
\mathrm{Im}
\left[
\frac{
\frac{-2\mu}{1+i\Gamma}  
	+ 2 g\rho_0 
}
{
	-\frac{4\mu^2}{1+\Gamma^2}
	+ 2 g\rho_0  \frac{4 i\Gamma \mu}{1+\Gamma^2}
}
\right]
=   \frac{c_{\text{J}}^2}{\mu} \Gamma \rho_0 
\frac{
\mu^2 + \omega_{\text{ex}}^2 + \frac{1}{16} \gamma^2
}
{
\mu^2+ \frac{1}{16}\gamma^2
}
\end{align}
where we have made use of the Green's function $G_\Phi=G_{11}$ from  \equa{eqn:green_function} and defined the damping rate  $\gamma=8g \rho_0 \Gamma$ with the unit of frequency, and $c_{\text{J}}$ should take the value of $c_{\text{J}}(\omega=\mu \approx \omega_{\text{ex}})$ in \equa{eqn:c_J_appendix}.

Note that the real part of \fig{fig:driving_diagram}(a) gives the time averaged Lagrangian of the dark condensate: 
\begin{align}
\langle L \rangle_t=c_{\text{J}}^2 \Phi^{\ast} \Phi \mathrm{Re}[G_\Phi(-2\mu)]
=c_{\text{J}}^2 \rho_0 \frac{1}{4}
 \frac{
-	(\mu + \omega_{\text{ex}}) - 2g \rho_0 \gamma^2
}
{
	\mu^2+ \frac{1}{16}\gamma^2
}
\xrightarrow{\gamma \rightarrow 0} 
c_{\text{J}}^2 \rho_0 \frac{1}{4}
\frac{
	-	(\mu + \omega_{\text{ex}}) 
}
{
	\mu^2
}
\label{eqn:mean_field_driving_energy_1}
\end{align}
which agrees with \equa{eqn:p_potential} in the zero damping limit.

\section{The static phase}
In the case $\omega_{\text{ex}}=G_0-V_{\text{g}}-E_{\text{b}}<0$ such that the condensate exists even without a bias $\mu$, the condensate may not be a dynamical one if the contact bias $\mu$ is very weak.
From \equa{eqn:H_drive} and the energy landscape in \fig{fig:Static_condensate}(d), the condensate  is a static and dark one at $\mu=0$ because its energy density at $\theta=\pm \frac{\pi}{2}$ is lower than the bright one by $c_{\text{J}} \rho_0$.
At nonzero $\mu$, the free energy landscape rotates according to \equa{eqn:H_drive}. Since we are discussing in the rotating frame, the system is in an actual dynamical state if the order parameter doesn't follow the rotation, and a static one if it follows it.
For a very small $\mu$,  the rotation of the free energy landscape is slow, and the order parameter tends to follow the instantaneous minimum adiabatically, meaning the system is in a static state.  At large enough $\mu$, the system couldn't catch up with the rotation of the landscape, and would be a dynamical state. In this section, we speculate the transition point $\mu_c$ from the static to the dynamical condensate.

In the regime of $|\omega_{\text{ex}}|\gg c_{\text{J}}$, i.e., the superfluid density is not too small, one can integrate out the amplitude to obtain the Lagrangian for the phase. After transforming back to the `lab frame', the phase Lagrangian reads
\begin{align}
	L=&  \frac{1}{4g} 
	\left[ 
	-\left(\dot{\theta}+\mu\right)^2
	- 2 c_{\text{J}} \omega_{\text{ex}} \cos (2 \theta)
	\right]
	\label{eqn:L_phase}
	\,.
\end{align}
It is obvious that as $\mu$ increases, the static-dynamic transition is the temporal analogue of the commensurate-incommensurate transition  \cite{Bak1982} in charge density wave systems, and the dynamical state just above the transition is a train of temporal solitons in time satisfying $\partial_t^2 \theta= -2 c_{\text{J}} \omega_{\text{ex}} \sin 2\theta$. Assuming a complete analogy with the commensurate-incommensurate transition in space, comparison of the average energy of this temporal soliton state with the static state says that the transition happens at $\mu =\mu_c= \frac{4}{\pi}\sqrt{-2 c_{\text{J}} \omega_{\text{ex}} }$ \cite{Bak1982} which is about $8 \unit{meV}$ for $-\omega_{\text{ex}}=0.1 \unit{eV}$ and the typical parameters used in this paper. However, since the transition is into a bright condensate as \fig{fig:vp_eta} whose phase action is no longer \equa{eqn:L_phase}, the actual $\mu_c$ is probably different although at the same order.
This transition may also be viewed as a transition to a time crystal that breaks the continuous time translational symmetry and develops long range order in the temporal  direction. We leave its details for future study.

\section{Collective modes}
\begin{figure}
	\includegraphics[width=0.6\linewidth]{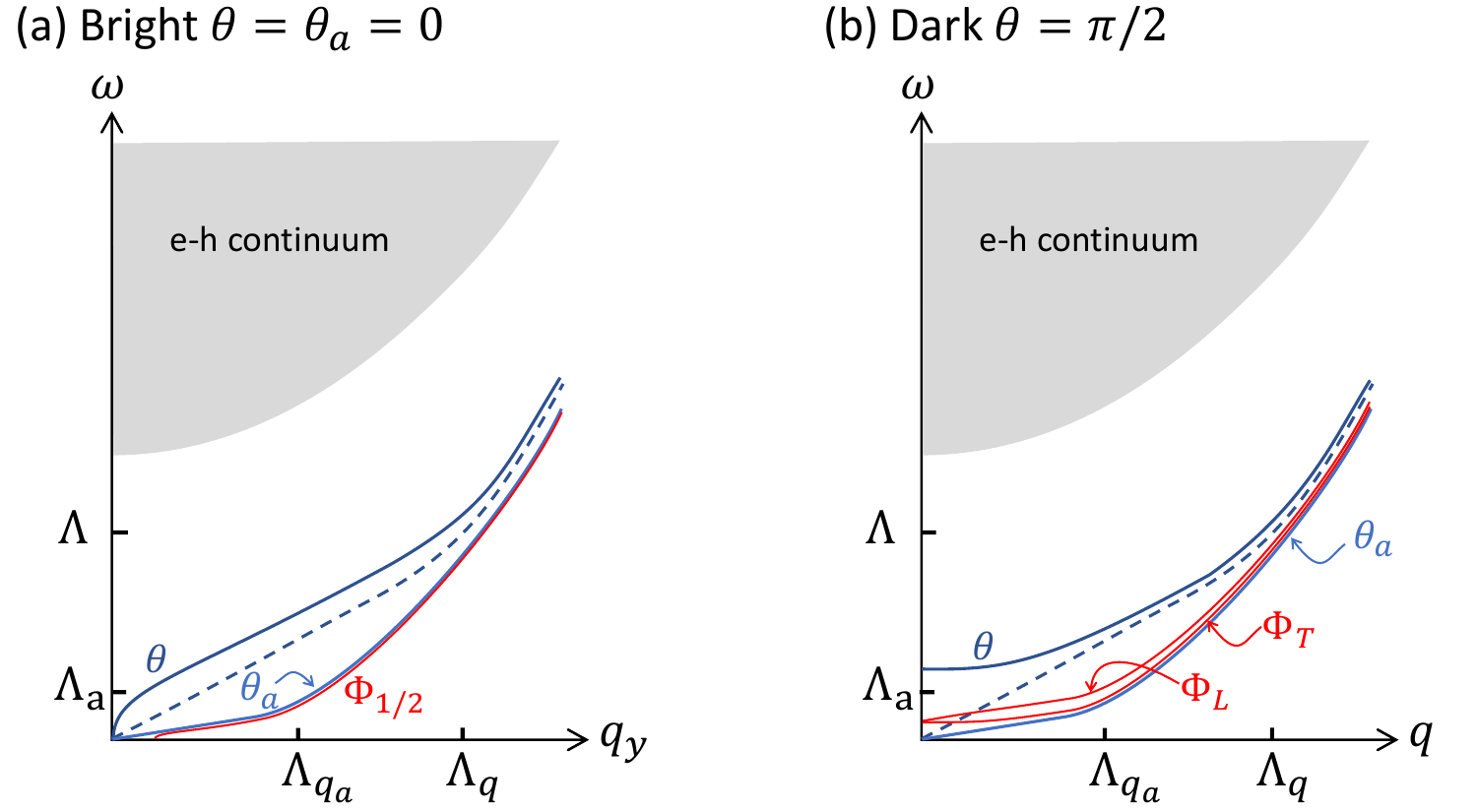} 
	\caption{(a) Instantaneous dispersion of the four phase modes of the bright condensate at a certain time in the rotating frame such that $\theta=\theta_0=0$. The fastest one is the sound mode (fluctuations of the overall phase $\theta$) and the other three are the valley pseudo-spin waves. At small momenta, they are the Goldstone modes modified by the $H_{\text{J}}$ term. At large momenta, they crossover to the dispersion of the single excitons. Note that  the dispersion of the modes are periodically modulated by the dynamical $H_J$ term.
		(b) Same as (a) but for the dark condensate. 
	}
	\label{fig:Collective_modes}
\end{figure}

In this section, we discuss the dispersion of the collective modes of the dynamical condensates~\cite{Wouters.2007} in the dissipationless limit ($\Gamma \rightarrow 0$).  On top of an arbitrary mean field order parameter, the collective modes are just the four Gaussian fluctuation modes expanded from \equa{eqn:H_drive}.

We start with the zero tunneling limit: $H_{\text{J}} =0$ which  essentially results in an equilibrium state. Around the state with only  $\Phi_0$ nonzero, the Lagrangian and dispersions for order parameter fluctuations $(\delta \Phi_0,  \Phi_i)$ are
\begin{align}
	L_{\Phi_0}=&
	\delta\Phi_0^\ast \left(-i \partial_t +  \frac{q^2}{4m} 
	+ 2g\rho  \right) 	\delta\Phi_0
	+ g  \left( \Phi_0^2 \delta\Phi_0^{\ast 2} +c.c.  \right)
	,\quad 
	\omega(q)^2=\frac{q^2}{4m} \left( \frac{q^2}{4m}+ 4g \rho \right)
	,\notag\\
	L_{\Phi_i}=&
	\Phi_i^\ast \left(-i \partial_t +  \frac{q^2}{4m}  \right) 	\Phi_i	
	+ \frac{c_4}{2} \left( \Phi_0^{\ast} \Phi_i  +c.c.  \right)^2
	,\quad 
	\omega_a(q)^2=\frac{q^2}{4m} \left( \frac{q^2}{4m}+ 2 c_4 \rho \right)
	\label{eqn:collective_mode_langrangian_0}
\end{align}
where $g=(U_0+c_4)/2$ and $i=1,2,3$. The same collective mode spectrum applies to any state in the $S_1 \times S_3$ ground state manifold since they can all be connected to the $\Phi_0$ by symmetry operations. 

With nonzero and dynamical $H_{\text{J}}$, the dispersion of the collective modes are periodically modulated by it. We can only describe their instantaneous dispersion  at each time.
It is now more convenient to represent the bright and dark condensates by
\begin{align}
	|\Phi| e^{i\theta}
	\begin{pmatrix}
		e^{i\theta_a} & 0
		\\
		0 & e^{-i\theta_a}
	\end{pmatrix}
	,\quad
	|\Phi| e^{i\theta}
	\begin{pmatrix}
		0 & e^{-i\theta_a}
		\\
		e^{i\theta_a} & 0
	\end{pmatrix}
\end{align}
which correspond to ($\Phi_0,\, \Phi_3 \neq 0$; $\Phi_1=\Phi_2 = 0$) and ($\Phi_0=\Phi_3 = 0$; $\Phi_1,\, \Phi_2 \neq 0$), respectively.

The dark state corresponds to $\Phi_\mu=e^{i\theta}(0, a_1, a_2, 0)$. 
The $c_{\text{J}}$ term reduces the U(1)$\times$U(1) invariance of varying $\theta$ and $\theta_a$ to U(1), such that the free energy looks like that in \fig{fig:Static_condensate}(b). After adding $H_J$ to \equa{eqn:collective_mode_langrangian_0} and integrating out the implicit EM fields, 
the dispersions of the four modes around $\theta_a=0$ are found to be
\begin{align}
	\omega_\theta(q)^2=&\omega(q)^2-\omega_J^2 \cos 2\theta
	,\quad
	\omega_{\theta_a}(q)^2=\omega_a(q)^2
	,\quad
	\omega_{\Phi_T}(q)^2 
	=\omega_a(q)^2
	-\frac{1}{2}\omega^2_{Ja} \cos 2\theta-2c_{\text{J}}  \frac{q^2}{4m}  \cos 2\theta
	,\notag \\
	\omega_{\Phi_L}(q)^2 
	=&\omega_a(q)^2
	-\frac{1}{2}\omega^2_{Ja} \cos 2\theta-2c_{\text{J}}  \frac{q^2}{4m}  \cos 2\theta
	+2c_{\text{em}}^2  \frac{2\pi q}{\epsilon} \left(\frac{q^2}{2m}+ c_4 ( 2\rho-\Phi_1^2-\Phi_1^{\ast 2})
	\right)
	\label{eqn:modes_inter_s}
\end{align}
to lowest order in $v_{\text{t}}^2$ where $\omega_{Ja}=2\sqrt{c_4 c_{\text{J}} \rho_0}$. The $\Phi_L, \, \Phi_T$ modes are hybridizations of $\Phi_0,\, \Phi_3$ fluctuations with `chiral' combination coefficients determined by the direction of momentum: $\Phi_L=\hat{q}_x \Phi_3 +  i\hat{q}_y \Phi_0,\, \Phi_T=-\hat{q}_y \Phi_3 +  i\hat{q}_x \Phi_0$, such that the dipole fluctuation of $\Phi_L$ is parallel to its momentum, while that of $\Phi_T$ is perpendicular. This physics is the same as LO-TO splitting of optically active excitons (A-exciton) in TMD monolayers~\cite{Yu2014}. 
Around the ground states $\theta=\pm \pi/2$ and at small momentum, the dispersion of the $\theta$ mode reduces to $\omega_q= \sqrt{v^2 q^2 + \omega_J^2}$, i.e., a Josephson plasmon~\cite{Sun2021a} where $\omega_J=2\sqrt{2g \rho_0 c_{\text{J}} }$ is the Josephson plasma frequency.  The dispersion of $\theta_a$ is unchanged by the Josephson coupling. We note that for typical parameters such that $G\approx E_{\text{b}}$ and $U_0 \gg c_4$, one has $\omega_J=4\sqrt{ \frac{1}{3}  m^2 v_{\text{t}}^2  v_0^2}$, in the THz range (see Table~\ref{tbl:typical values}).

The bright condensate corresponds to $\Phi_\mu=e^{i\theta}(i a_0, 0 , 0, a_3)$. In the intravalley state, the symmetry of varying $\theta$ is accidentally preserved in our leading order approximation to the tunneling function in \equa{eqn:tk}, such that the energy is still $U(1)$ invariant on the complex plane, as shown in \fig{fig:Static_condensate}(a). 
The dispersions of the four  modes are found to be
\begin{align}
	\omega_{\theta}(q)^2 &=\omega(q)^2
	+ 2D \frac{\cos^2 (\theta)}{q \epsilon} \left(q_y \cos\theta_a+q_x \sin\theta_a \right)^2
	,\notag\\
	\omega_{\theta_a}(q)^2 &=\omega_a(q)^2+ 2D_a\frac{\sin^2 (\theta)}{q \epsilon} \left(q_y \sin\theta_a-q_x \cos\theta_a \right)^2
	\,,\quad
	\omega_{\Phi_j}(q)^2 
	=\omega_a(q)^2- \frac{1}{2} \omega_{Ja}^2 \cos 2\theta
	\label{eqn:modes_intra_s}
\end{align} 
to the lowest order in $v_{\text{t}}^2$. Note that we have neglected the Coulomb cross coupling ($\sim v_{\text{t}}^2$) between the $\theta$ and $\theta_a$ modes, which is negligible at large momenta but important at small momenta. The $\Phi_j$ modes are stable or unstable depending on the overall phase $\theta$.
It is obvious that the effect of $\theta_a$ is to rotate the dispersion of the $\theta,\, \theta_a$ modes on the x-y plane by $\theta_a$, while it does not affect the $\Phi_1, \Phi_2$ modes. Due to the locking between the phase and polarization, certain phase fluctuations are accompanied by local charge density fluctuations, and therefore experience Coulomb renormalization such that they are plasmon-like at small momenta. 	For example, the $\theta$ mode (exciton sound wave) dispersion is $\omega_q= \sqrt{v^2 q^2 + 2D q_y^2/(\epsilon q)}$ at small momenta where  $D=8\pi (c_4+U_0) P_0^2$ is a Drude weight. For $U_0 \gg c_4$, it is about $D=\frac{2^7 \pi}{25} \frac{m v_{\text{t}}^2}{e^2/d} \frac{\rho e^2}{m}$. As shown by the schematic dispersion along $q_y$ in \fig{fig:Collective_modes}(a), it crossovers from $\omega \propto \sqrt{q}$ to $\omega \approx v q$ at the momentum scale $q_c \sim \frac{64}{25} mv_{\text{t}}^2/(\epsilon e^2)$ which is about $2\pi/(130 \unit{\mu m})$ for typical parameters.

\section{Single species state}
If the exchange interaction $c_4$ is negative, the condensate prefers the single species state where only one component in $\Phi_{ij}$ is nonzero with the exciton density $\rho=(-\omega_{\text{ex}}-c_{\text{J}} v_{\text{t}}^2)/(2c_4+U_0)$, and the ground state manifold has only one phase degree of freedom. This happens when the interlayer distance is large compared to the Bohr radius \cite{Wu2015,Combescot2015}.

In the intravalley state represented by, e.g., a nonzero $\Phi_{11}=\sqrt{\rho}e^{i\theta}$, the in plane polarization is $\mathbf{P}=P_0\left(\cos \theta \hat{x} +\sin \theta \hat{y} \right)$ whose direction is locked with the phase. In the intervalley state represented by, e.g., a nonzero $\Phi_{12}=\sqrt{\rho}e^{i\theta}$, there is no in plane polarization.

\section{First order Josephson coupling}
\label{sec:first_order}
In a class of bilayers, the interlayer tunneling $t_{\text{k}}=t_0$ does not depend on momentum, leading to the  scenario of first order Josephson effect. The Josephson coupling  in the rotating frame \equa{eqn:H_drive} is  changed to
\begin{align}
H_{\text{J}}
= & 
\int dr \left[
c_{t} t_0 \Phi_0 e^{-i\mu t} + c_{t^2} t_0^2 \left( 
\rho +\Phi_\mu \Phi^\mu e^{-2i\mu t}  \right)
+c.c. \right]
\,.
\label{eqn:Lagrangian_first_order}
\end{align}
The coefficient $c_t$ is shown in \fig{fig:driving_diagram}(b), which reads
\begin{align}
c_t &=  \chi_{t_0,\Phi}(0)
= \sum_k \varphi_k
\frac{\omega_{\text{ex}}-2\xi_k}{-2\xi_k} 
=  \frac{ \sqrt{8\pi}}{\sqrt{m}} \sum_k \frac{E_{\text{b}}}{(E_{\text{b}}+2\varepsilon_k)^{1/2}(G+2\varepsilon_k)}
\notag\\
&=
\left\{
\begin{array}{lc}
\sqrt{\frac{2 m}{\pi}}   E_{\text{b}}  \frac{\cos^{-1}\sqrt{\frac{E_{\text{b}}}{G}}}{\sqrt{G-E_{\text{b}}}}
\xrightarrow{G\gg E_{\text{b}}}
\sqrt{\frac{\pi m}{2}}    \frac{E_{\text{b}}}{\sqrt{G}}
= \frac{\sqrt{2\pi}  }{a_0} \sqrt{\frac{E_{\text{b}}}{G}} \,,
& G>E_{\text{b}}
\\
\sqrt{\frac{2 m}{\pi}}   E_{\text{b}}  \frac{\cosh	^{-1}\sqrt{\frac{E_{\text{b}}}{G}}}{\sqrt{E_{\text{b}}-G}}
\xrightarrow{G \approx E_{\text{b}}}
\sqrt{\frac{\pi m}{2}} \sqrt{E_{\text{b}}}=   \frac{\sqrt{2\pi}  }{a_0} \,,
& G<E_{\text{b}}
\end{array}
\right..
\end{align}

In the non equilibrium case $\mu \neq 0$, the $c_t$ term linearly drives the order parameter $\Phi_0$. Setting $\Gamma=0$ and integrating out this `fast' degree of freedom gives a  $O(t_0^2)$ driving energy  through linear response (\fig{fig:driving_diagram}(c)):
\begin{align}
 V_{\text{singlet}} =c_t^2 t^2_0 2 \frac{-\mu+(c_4+U_0)\rho}{\mu^2} 
=c_t^2 t^2_0 2 \frac{-\omega_{\text{ex}}}{\mu^2} 
\,,\quad
V_{\text{triplet}} =c_t^2 t^2_0 2 \frac{-\mu+c_4 \rho}{\mu^2}
\,
\end{align}
for the valley-singlet ($\Phi_0$) and the valley-triplet  ($\Phi_1,\Phi_2,\Phi_3$) condensates, respectively.

We now speculate the phase diagram in the dissipationless limit. As shown in Ref.~\cite{Sun.2023_ponderomotive}, without dissipation, this driving energy is what the steady state minimizes.
Since $V_{\text{singlet}} >V_{\text{triplet}} $,  the dynamical condensate would be the valley-triplet one. However, for $\mu<\mu_c \sim g c_t t_0 \sqrt{\rho_0}$, the system should be  the static singlet condensate.

If the tunneling does not conserve  the in plane momentum and valley index, the first order Josephson term becomes $c_{t} t_q \sum_q(\Phi_0+\Phi_1)_q + c.c.$ with the `disordered' tunneling matrix element satisfying $\langle t_q t_{-q}\rangle = f(q)$. This may be the case of two TMD monolayers separated by  incommensurate spacers, which we leave for future study.

\section{Typical parameters}
The typical parameters for the TMDB  bilayer devices are shown in Table~\ref{tbl:typical values} for the case of second order Josephson coupling.

\begin{table}
	\begin{ruledtabular}
		\begin{tabular}{lll}
			Symbol & Physical Meaning  & Value
			\\
			\hline
			$E_{\text{b}}$ & binding energy & $271 \unit{meV}$
			\\[3pt]
			$a_0$ & exciton size & $2.1 \unit{nm}$
			\\[3pt]
			$\rho$ & exciton density & $5.1 \times 10^{11} \unit{cm^{-2}}$
			\\[3pt]
			$v$ &  $\theta$ phase mode (sound) speed& $6.6 \times 10^{4} \unit{m/s}$
			\\[3pt]
			$v_a$ & valley-pseudo spin wave speed &$5.8 \times 10^{4} \unit{m/s}$
			\\[3pt]
			\hline
			& \textbf{Bright Condensate}  &
			\\[3pt]
			$E$ & Radiated electric field &$170 \unit{V/cm}$
			\\[3pt]
			$P_{\text{r}}$ & Radiation power     &$0.8 \unit{\mu W}$
			\\[3pt]
			$I_{\text{bright}}$ & Tunneling current    &$\sim 1 \unit{\mu A}$
			\\[3pt]
			$P_0$ & Polarization density & $\frac{2/5}{(2\pi)^{1/2}} \frac{\hbar v_{\text{t}}}{e^2} \sqrt{\rho_0}$
			\\[3pt]
			$D$ & Drude Weight for the phase `plasmon'  & $\frac{2^7 \pi}{25} \frac{m v_{\text{t}}^2}{e^2/d} \frac{\rho_0 e^2}{m}$
			\\[3pt]
			\hline
			& \textbf{Dark Condensate}  &            
			\\[3pt]
			$J_c$ & Josephson critical current &$0.9 \unit{mA}$ 
			\\[3pt]
			$\omega_J$ & Josephson plasma frequency &$4.4 \unit{meV}$
			\\[3pt]
			$I_{\text{dark}} $ & Tunneling current  & $\sim 0.1 \unit{\mu A}$
		\end{tabular}
	\end{ruledtabular}
	\caption{Values of physical quantities using the parameters for WSe$_2$/MoSe$_2$ bilayer \cite{Ma.2021strongly} without an hBN spacer or a cavity: $m=0.5 m_e$, $G_0=1.5 \unit{eV}$, $G=G_0-V_{\text{g}}=1 \unit{eV}$, $\mu=0.754 \unit{eV}$, $\epsilon=10$, $d=0.6 \unit{nm}$,  and $v_{\text{t}} =10^4 \unit{m/s}$. The device size is $1 \unit{\mu m^2}$. To compute $I_{\text{dark}}$, the  device-lead tunneling rate is set to $\gamma = 0.1 \unit{eV}$. }
	\label{tbl:typical values}
\end{table}

\end{document}